%% file: main.tex
\documentclass[12pt]{article}

\newcommand{\blind}{0}

\usepackage{amsmath}
\usepackage{graphicx,psfrag,epsf}
\usepackage{enumerate}
\usepackage{natbib}
\usepackage{url} 
\usepackage{amsfonts}

\usepackage[a4paper]{geometry}

\usepackage[latin1]{inputenc}
\usepackage{natbib}

\usepackage[english]{babel}

\usepackage{tabularx}
\usepackage{rotating}
\usepackage{xspace}
\usepackage{bm}
\usepackage[ruled,vlined]{algorithm2e}

\usepackage[pdftex,dvipsnames]{xcolor}
\usepackage{xargs}  

\usepackage{tikz}
\usepackage{subcaption}

\tikzset{
    cross/.pic = {
    \draw[rotate = 45] (-#1,0) -- (#1,0);
    \draw[rotate = 45] (0,-#1) -- (0, #1);
    }
}
\definecolor{salmon}{RGB}{205,112,84}
\DeclareRobustCommand{\dotdash}{\tikz[baseline=-0.6ex]\draw [thick,dashdotted]{(0,0) -- (0.5,0)};}
\DeclareRobustCommand{\solid}{\tikz[baseline=-0.6ex]\draw [thick,solid]{(0,0) -- (0.5,0)};}
\DeclareRobustCommand{\dotted}{\tikz[baseline=-0.6ex] \draw[thick, dotted](0,0)--(0.5,0);}
\DeclareRobustCommand{\dashed}{\tikz[baseline=-0.6ex]\draw [thick,dashed]{(0,0) -- (0.5,0)};}
\DeclareRobustCommand{\vsolid}{\enspace\tikz[baseline=0ex]\draw [thick,color = salmon,solid]{(0,-0.1) -- (0,0.3)};\enspace}
\DeclareRobustCommand{\cdashed}{\tikz[baseline=0.2ex]\draw [thick,color = salmon, solid]{(0,0) -- (0.5,0.3)};}
\DeclareRobustCommand{\cross}{\tikz[baseline=-0.7ex]\path (0,0) pic[thick,color = salmon] {cross=4pt};}
\DeclareRobustCommand{\square}{\tikz[baseline=-0ex]\fill [salmon,opacity=0.7] (0,0) rectangle (0.25,0.25);}
\definecolor{col1}{HTML}{F8766D}
\definecolor{col2}{HTML}{7CAE00}
\definecolor{col3}{HTML}{00BFC4}
\definecolor{col4}{HTML}{C77CFF}

\usepackage[colorinlistoftodos]{todonotes}
\newcommandx{\improvement}[2][1=]{\todo[linecolor=Plum,backgroundcolor=Plum!25,bordercolor=Plum,#1]{#2}}

\newcommand\code{\bgroup\@codex}
\def\@codex#1{\small {\normalfont\ttfamily\hyphenchar\font=45 #1}\egroup}

\include{martin-math-commands}

\begin{document}


\def\spacingset#1{\renewcommand{\baselinestretch}%
{#1}\small\normalsize} \spacingset{1}


\if0\blind
{
  \title{\bf Importance Sampling with the Integrated Nested Laplace Approximation}
  \author{Martin Outzen Berild 1\thanks{
    V. G\'omez-Rubio has been supported by grant SBPLY/17/180501/000491, funded by
Consejer\'ia de Educaci\'on, Cultura y Deportes (JCCM, Spain) and FEDER, and
grants MTM2016-77501-P and PID2019-106341GB-I00, funded by Ministerio de Ciencia e Innovaci\'on (Spain).}\hspace{.2cm}\\
    Department of Mathematics,\\ Norwegian University of Science and Technology, Norway\\
    and \\
    Sara Martino \\
    Department of Mathematics,\\ Norwegian University of Science and Technology, Norway\\
    and \\
    Virgilio G\'omez-Rubio\\
    Department of Mathematics,\\
    School of Industrial Engineering-Albacete,\\ Universidad de Castilla-La Mancha, Spain\\
    and \\
    H\aa{}vard Rue\\
    CEMSE Division, King Abdullah University of Science and Technology,\\
    Thuwal 23955-6900, Saudi Arabia}
  \maketitle
} \fi

\if1\blind
{
  \bigskip
  \bigskip
  \bigskip
  \begin{center}
    {\LARGE\bf Importance Sampling with the Integrate Nested Laplace Approximation}
\end{center}
  \medskip
} \fi

\bigskip
\begin{abstract}
The Integrated Nested Laplace Approximation (INLA) is a deterministic approach to Bayesian inference on latent Gaussian models (LGMs) and focuses on fast and accurate approximation of posterior marginals for the parameters in the models. 
Recently, methods have been developed to extend this class of models to those that can be expressed as conditional LGMs by fixing some of the parameters in the models to descriptive values. These methods differ in the manner descriptive values are chosen. This paper proposes to combine importance sampling with INLA (IS-INLA), and extends this approach with the more robust adaptive multiple importance sampling algorithm combined with INLA (AMIS-INLA).

This paper gives a comparison between these approaches and existing methods on a series of applications with simulated and observed datasets and evaluates their performance based on accuracy, efficiency, and robustness. The approaches are validated by exact posteriors in a simple bivariate linear model; then, they are applied to a Bayesian lasso model, a Bayesian imputation of missing covariate values, and lastly, in parametric Bayesian quantile regression. The applications show that the AMIS-INLA approach, in general, outperforms the other methods, but the IS-INLA algorithm could be considered for faster inference when good proposals are available.  \end{abstract}

\noindent%
\vfill

\newpage
\spacingset{1.5} 

\section{Introduction}
\label{sec:intro}

The integrated nested Laplace approximation \citep[INLA,][]{isi:000264374200002} is a numerical method for approximated Bayesian inference on a well determined class of models named Latent Gaussian models (LGMs). INLA focuses on providing approximate marginal posterior distributions for all parameters in the model. This is in contrast with the more traditional Markov Chain Monte Carlo \citep[MCMC,][]{Gilksetal:1996} based inference that provides instead an estimate of the join posterior distribution. INLA has become a widely used method because it is, in general, faster than MCMC while still providing accurate estimates. Moreover, INLA is implemented as an R package called R-INLA, that allows the user  to do inference on complex hierarchical models often in a matter of seconds. 

Implementing INLA from scratch may be a difficult task, therefore, fitting models with INLA is, in practice, restricted to the classes of models implemented in the R-INLA package. How to enlarge such selection has been the topic of many papers \citep[see, for example,][and the references therein]{Bivandetal:2014,Bivandetal:2015,GomezRubioetal:2020}. One interesting approach is the one taken in \cite{GomezRubioRue:2018} where they propose to combine INLA and MCMC methods. The basic idea is that certain models, named  conditional LGMs, can be fitted with  INLA, provided  a (small) number of parameters are fixed to a given value. 
\cite{GomezRubioRue:2018} propose to draw samples from the posterior distribution of the conditioning parameters by combining MCMC techniques and  conditional models fitted with R-INLA. This is made possible by the fact that INLA computes also the marginal likelihood of the conditional fitted model. The marginal likelihood is used, in \cite{GomezRubioRue:2018} to compute the acceptance probability in the Metropolis-Hastings (MH) algorithm, which is a popular MCMC method. 

Combining INLA and MCMC  allows to increase the number of models that can be fitted using R-INLA. The MCMC algorithm is simple to implement as only the conditioning parameters need to be sampled while the  rest of the parameters are integrated out using INLA. The INLA-MCMC approach proposed by  \cite{GomezRubioRue:2018} relies on the MH algorithm and requires model fitting with R-INLA at every step. That may be slow
in practice because the sequential nature of the MH algorithm makes parallelization. \cite{GomezRubioPalmiPerales:2019} provide some insight
on how to speed up the process of fitting conditional models with INLA, but
it requires a good approximation to the posterior mode of the parameters
of interest by relying, for example, on maximum likelihood estimates. 

In this paper we propose a new method for model fitting with INLA, similar in spirit to  \cite{GomezRubioRue:2018} but based on the importance sampling (IS) algorithm instead of on the MH one. The big advantage of the IS algorithm over MH is that it is easy to parallelize, thus allowing for a great improve in computational speed. The drawback is that, lacking the adaptive nature of the MH algorithm, the performance of IS based inference relies on the choice of  a good proposal distribution. This can be hard to determine in many practical cases. We propose therefore also an algorithm that is based on an adaptive multiple IS \citep{corneut_adaptive_2012} that, for a slightly higher computing time, has the advantage of requiring less human intervention. 

The rest of the paper is organized as follows. The class of models amenable to INLA are described in Section \ref{sec:INLA}. A short description of how INLA works is also given in the same Section. Section \ref{sec:IS} introduces importance sampling while Section \ref{sec:ISINLA} shows how INLA and IS can be combined. In this section we also discuss numerical and graphical diagnostic to assess the accuracy of our algorithm.  In Section \ref{sec:amis} an adaptive version of the algorithm is presented while in Section \ref{sec:examples} we show, in several examples, how our prosal works in practive. We end with a discussion in Section \ref{sec:discussion}.

\section{The Integrated Nested Laplace Approximation}
\label{sec:INLA}
 Let our response $\bm{y} = (y_1, \dots, y_n)$ form a vector of observations from a distribution in the exponential family with mean $\mu_i$. We assume that a linear predictor $\eta_i$, can be related to $\mu_i$ using an appropriate link function:
\begin{equation}\label{eq:linpred}
    \eta_i = g(\mu_i) = \alpha  + \sum_{k=1}^{n_\beta} \beta_k z_{ki} + \sum_{j=1}^{n_f} f^{(j)}(u_{ji}) + \epsilon_i
\end{equation}
The  predictor consists of linear terms on some covariates $\bm{z}_k$, and some other terms such as random effects, spatial effect, non-linear effects of the covariates, etc., defined by some indices $\bm{u}_j$. All these terms define a latent field $\bm{x} = (\bf{\eta},\alpha, \bm{\beta},{\bm f}^{(1)},{\bm f}^{(2)},\dots)$. The likelihood and the prior for $\bm{x}$ will depend on some hyperparameters $\bm{\theta}$ and an appropriate prior $\pi(\bm{\theta})$ is assigned to these. 

From Equation \eqref{eq:linpred}, it is clear that the observations are conditionally independent given the latent effect $\bm{x}$ and the hyperparameters $\bm{\theta}$ so that the likelihood can be written as 
\begin{equation}
    \pi(\bm{y}|\bm{x},\bm{\theta}) = \prod_{i\in\mathcal{I}} \pi(y_i|x_i,\bm{\theta}),
    \label{eq:INLA-Like}
\end{equation}
where $i$ belongs to a set $\mathcal{I} = (1,\dots,n)$ that indicates observed responses.

In a Bayesian framework, the main interest lays in the posterior distribution:
\begin{equation}\label{eq:posterior}
\pi(\bm{x},\bm{\theta}|\bm{y}) \propto \pi(\bm{x}|\bm{\theta})\pi(\bm{\theta})\prod_{i\in\mathcal{I}} \pi(y_i|x_i,\bm{\theta})
\end{equation}
This is usually not available in closed form, thus several estimation methods and approximations have been developed. 
INLA, introduced by \cite{isi:000264374200002}, is one of such methods. INLA can be used for LGM provided the prior for the latent field $\bm{x}$ is a Gaussian Markov random field (GMRF) model \citep{RueHeld:2005}. We assume the latent GMRF to have 0 mean and precision (inverse of covariance) matrix $\bm{Q(\theta)}$. Equation~\eqref{eq:posterior} can then be rewritten as 
 \begin{equation}
    \pi(\bm{x},\bm{\theta}|\bm{y}) \propto \pi(\bm{\theta})|\mathbf{Q}(\bm{\theta})|^{1/2}\exp\left\{-\frac{1}{2}\bm{x}^T
    \mathbf{Q}(\bm{\theta})\bm{x} + \sum_{i\in\mathcal{I}} \ln(\pi(y_i|x_i,\bm{\theta})) \right\}
    \label{eq:INLA-post}
\end{equation}

INLA does not seek to approximate the joint posterior distribution $ \pi(\bm{x},\bm{\theta}|\bm{y})$, instead, it creates numerical approximations to the posterior marginals for the latent field $\pi(x_i|\bm{y})$ and the hyperparameters $\pi(\theta_j|\bm{y})$. To do this, the first step is to approximate $ \pi(\bm{\theta}|\bm{y})$ by $ \tilde{\pi}(\bm{\theta}|\bm{y})$. Approximated marginal posteriors for the hyperparameters $\tilde{\pi}(\theta_j|\bm{y})$ can then be derived from $\tilde{\pi}(\bm{\theta}|\bm{y})$ via numerical integration. Posterior marginals for the latent field $\pi(x_i|\bm{y})$ can be written as 
\begin{equation}
\pi(x_i|\bm{y}) = \int \pi(x_i|\bm{\theta},\bm{y})\pi(\bm{\theta}|\bm{y}) d\bm{\theta} 
\end{equation}
and approximated as 
\begin{equation}
    \tilde{\pi}(x_i|\bm{y}) = \sum_{g} \tilde{\pi}(x_i|\theta_g,\bm{y})|\tilde{\pi}(\theta_g|\bm{y}) \Delta_g
    \label{eq:INLA-post-m}
\end{equation}
where $\theta_g$ are selected points and $\tilde{\pi}(x_i|\theta_g,\bm{y})$ is an approximation to $\pi(x_i|\theta_g,\bm{y})$, see \cite{isi:000264374200002} for details.

As a by-product of the main computations, INLA provides  other quantities of interest. Of importance for this paper is the marginal likelihood $\pi(\bm{y})$, which can be computed as: 
\begin{equation}
    \tilde{\pi}(\bm{y}) = \int \frac{\pi(\bm{y}|\bm{x},\bm{\theta})\pi(\bm{x}|\bm{\theta})
    \pi(\bm{\theta})}{\tilde{\pi}_G(\bm{x}|\bm{\theta},\bm{y})} 
    \bigg|_{\bm{x} = \bm{x}_0(\bm{\theta})} \dd \bm{\theta}
    \label{eq:INLA-m-Like}
\end{equation}
Here $\tilde{\pi}_G(\bm{x}|\bm{\theta},\bm{y})$ is a Gaussian approximation of $\pi(\bm{x}|\bm{\theta},\bm{y})$ build by matching the mode and the curvature at the mode and $\bm{x}_0(\bm{\theta})$ is the posterior mode of $\bm{x}|\bm{\theta}$.  \cite{HubinStorvik:2016b} have investigated the performance of this approximation, finding it very accurate for a large class of models. Several authors \citep{Bivandetal:2014,Bivandetal:2015, GomezRubioRue:2018,GomezRubioPalmiPerales:2019,GomezRubioetal:2020} have relied on the estimates of the marginal likelihood provided by INLA for model estimation and they have found them to be accurate enough in a number of scenarios.

\section{Importance Sampling}
\label{sec:IS}

Importance sampling (IS) is a popular Monte Carlo method where a mathematical expectation with respect to a target distribution is approximated by a weighted average of random draws from another distribution. 
IS relies on a simple probability result, which is stated next.

Let $\pi(x)$ be a probability density function for the random variable $X$ defined on $\mathcal{D}\subseteq \reals^d$,  $d\geq 1$, and assume that we wish to compute $\mu_{\pi}$ defined as 
\begin{equation}
    \mu_{\pi}  = \mathbb{E}_{\pi}[h(X)] = 
    \int_\mathcal{D} h(x) \pi(x) \dd x
    \label{eq:MC}
\end{equation}
where $h(\cdot)$ is some function of $X$. Then for any probability density $g(x)$ that satisfies $g(x)>0$ whenever $h(x)\pi(x)>0$, it holds that
\begin{equation}
    \mu_{\pi} = \mathbb{E}_{g}[h(X)w(X)]
\end{equation}
where the $w(x) = \frac{f(x)}{g(x)}$ and $\mathbb{E}_{g}[\cdot]$ indicates the expectation with respect to $g(x)$. Independent draws $\left\{x^{(j)}\right\}_{j = 1}^N$ from $g(x)$ can then be used to approximate $\mu_{\pi}$ as
\begin{equation}
    \hat{\mu}_{IS} = \frac{1}{N}\sum_{i = 1}^N h(x_i) w(x_i) 
    \label{eq:ISestimate}
\end{equation}

In many cases $\pi(x)$ is only known up to a normalizing constant, in these cases
$\hat{\mu}_{IS}$ is replaced by 
\begin{equation}
    \tilde{\mu}_{IS} =  \sum_{i=1}^N  h(x_i) \bar{w}(x_i)
    \label{eq:ISestimate2}
\end{equation}
where the so called self normalizing weights 
\begin{equation}
    \bar{w}(x_i) = \frac{w(x_i)}{\sum_{i=1}^N w(x_i)},
    \label{eq:ISweight}
\end{equation} 
can be computed as the  normalizing constant cancels out. The estimator based on the self normalizing weights is slightly biased but tend to improve the variance of estimates \citep{robert_monte_2004}.

The performance of the IS estimator, both in its original and self-normalizing form, depends on the choice of the proposal distribution $g(\cdot)$, which should be as close as possible to $\pi(\cdot)$. In fact, an improper choice, e.g. lighter tails in $g(\cdot)$, might lead to unbounded weights such that estimates only relies on few samples. 

A common measure of the efficiency of the algorithm is the effective sample size (ESS). An estimate  can be easily computed as
\begin{equation}
    \widehat{\mathrm{ESS}} = \frac{\left(\sum_{i=1}^n w_i\right)^2}{\sum_{i=1}^n w_i^2}
    \label{eq:ess}
\end{equation} 

This quantity is useful to assess the correlation of the simulated data
and provides an overall estimate of the amount of data obtained with sampling. However, effective sample size and estimation error are further discussed in Section~\ref{subsec:error}.

\section{Importance Sampling with INLA}
\label{sec:ISINLA}

In this Section we discuss how the class of models that INLA can fit can be extended by combining INLA and IS. Our approach follows the path presented  in \cite{GomezRubioRue:2018} with the key difference that we use IS instead of the MH algorithm.

Similar to \cite{GomezRubioRue:2018} we collect all unknown parameters of the model in the vector $\bm{z} = (\bm{x}, \bm{\theta})$ which is split into two subsets $\bm{z} = (\bm{z}_{-c},
\bm{z}_c)$, where $\bm{z}_{-c}$ indicates all parameters in $\bm{z}$ that are not included in $\bm{z}_c$. The vectors $\bm{z}_c$ and $\bm{z}_{-c}$ are chosen such that the posterior distribution of $\bm{z}$ can be written as
\begin{equation}
    \pi(\bm{z}|\bm{y})\propto  \pi(\bm{y}|\bm{z}_{-c},\bm{z}_{c}) \pi(\bm{z}_{-c}|\bm{z}_{c})\pi(\bm{z}_{c}).
    \label{eq:postisinla1}
\end{equation}
Furthermore, we assume that this model cannot be fitted with R-INLA unless the parameters in $\bm{z}_c$ are fixed to some appropriate values, i.e. we model $\bm{z}_{-c}$ given $\bm{z}_c$. Conditional on $\bm{z}_c$,  R-INLA can produce approximations to  the conditional posterior marginals $\pi(z_{-c,k}|\bm{y},\bm{z}_c)$ , where $k$ indicates the $k$th element of $\bm{z}_{-c}$, and to the conditional marginal likelihood $\pi(\bm{y}|\bm{z}_c)$, using Equations~\eqref{eq:INLA-m-Like} and 
\eqref{eq:INLA-post-m} respectively.

Unconditional posterior marginal for the elements of $\bm{z}_{-c}$ could then be obtained  integrating over $\bm{z}_c$ as
\begin{equation}
\pi(z_{-c,k}|\bm{y}) = 
\int \pi(z_{-c,k}, \bm{z}_c|\bm{y}) d\bm{z}_c =
\int \pi(z_{-c,k} | \bm{y}, \bm{z}_c) \pi(\bm{z}_c|\bm{y}) \dd\bm{z}_c.
\label{eq:ismarg}
\end{equation}
Here, the conditional posterior marginals $\pi(z_{-c,k} | \bm{y}, \bm{z}_c)$ are approximated with R-INLA. 

A na\"ive Monte Carlo estimate of the integral in Equation~\eqref{eq:ismarg} is 
 not a viable option; however, IS could be used to sample from a raw approximation $g(\bm{z}_c)$ of $\pi(\bm{z}_c|\bm{y})$, the posterior marginal in Equation~\eqref{eq:ismarg} can be approximated as
\begin{equation}
\tilde{\pi}(z_{-c,k}|\bm{y}) \simeq
\sum_{j=1}^n w_j \tilde{\pi}(z_{-c,k} | \bm{y}, \bm{z}_c^{(j)})
\label{eq:margIS}
\end{equation}
where $\bm{z}_c^{(j)}$ are samples from a (multivariate) sampling distribution $g(\cdot)$,  $\tilde{\pi}(z_{-c,k} | \bm{y}, \bm{z}_c^{(j)})$ are the approximated condition posterior marginals obtained by INLA and $w_j$ are the posterior weights defined as:
\begin{equation}
w_j \propto  \frac{\pi(\bm{z}^{(j)}_c|\bm{y})}{g(\bm{z}^{(j)}_c)} \propto \frac{\pi(\bm{y}|\bm{z}^{(j)}_c) \pi(\bm{z}^{(j)}_c)}{g(\bm{z}^{(j)}_c)}
\label{eq:ISweights}
\end{equation}

Note that we use the self normalizing version of the IS algorithm as in Equation~\eqref{eq:ISweights}.
In computing $w_j$ we need the conditional marginal likelihood $\pi(\bm{y}|\bm{z}^{(j)}_c)$ which, conveniently, is one of the outputs from R-INLA. See Section \ref{subsec:error} for a discussion on this.

Lastly, the joint posterior distribution of $\bm{z}_c$ can be found with 
\begin{equation}
    \pi(\bm{z}_c | \bm{y}) = \sum_{j=1}^n w_j \delta(\bm{z}_c-\bm{z}^{(j)}_c),
    \label{eq:kde}
\end{equation}
where $\delta(\cdot)$ is the Dirac delta function. This has also been noted in \cite{Elviraetal:2018}. In a practical manner, as Equation~\eqref{eq:kde} would require $n\rightarrow \infty$, the joint posterior distribution, $\pi(\bm{z}_c | \bm{y})$, is approximated using a weighted non-parametric kernel density estimation \citep{venables.ripley:modern}. A similar approach is used to find its posterior marginals $\pi(z_{c,k} | \bm{y})$ for the $k$-th element of $\bm{z}_c$.

\subsection{Choice of the sampling distribution}
\label{subsec:sampdist}
 
The sampling distribution $g(\bm{z}_c)$ needs to be chosen with care in order to 
have a good performance of the IS algorithm. In principle, it should be as close
as possible to $\pi(\bm{z}_c | \bm{y})$ but this may be difficult in practice.

We assume that $\bm{z}_c$ is a vector of real valued parameters (transformations might be applied if necessary), therefore
 $g(\bm{z}_c)$ is a multivariate distribution. A reasonable proposal could be a multivariate Gaussian or Student-$t$ with $\nu$ degrees of freedom. We indicate the location and scale parameters of both the Gaussian and Student-$t$ as  $\bm{\lambda} = (\bm{\mu},\bm{\Sigma})$. In the Student-$t$ case, for $\nu>2$ the covariance is defined as $\frac{\nu}{\nu - 2}\Sigma$. 
  We want to choose $\bm{\lambda}$ such that the proposal  is close to the target distribution. Moreover, for the Student-$t$ we want $\nu$ to be low to guarantee heavy tails. We start therefore  from a preliminary proposal $g_0(\bm{z}_c)$, with parameters $\bm{\lambda}_0 = (\bm{\mu}_{0},\bm{\Sigma}_{0})$.   Then, $N_0$ samples are generated from $g_0(\bm{z}_c)$ and used to build a rough approximation of the location  and scale of the target as: 
\begin{align}
    \bm{\mu}_1 & = \sum_{j = 1}^{N_0}\bar{w}^{(j)}\bm{z}_c^{(j)}\\
    \bm{\Sigma}_1 & = \sum_{j=0}^{N_0} \bar{w}^{(j)}(\bm{z}_c^{(j)} - \bm{\mu}_1)(\bm{z}_c^{(j)} - \bm{\mu}_1)^{\top},
    \label{eq:iscov}
\end{align}
\noindent
where $\bm{z}_c^{(j)}\sim g_0(\bm{z}_c)$ and  $\bar{w}^{(j)}$ is the normalized importance weight of the $j$th sample calculated with Equation~\eqref{eq:ISweight}.

The initial $N_0$ samples are then discarded and the new (improved) proposal distribution has parameters $\bm{\lambda}_1 = (\bm{\mu}_{1},\bm{\Sigma}_{1})$.  Other distributions than the Gaussian and the Student-$t$ could be used. For example,
correction for skewness could be included in the previous approach or
distributions with fatter tails could be employed.

\subsection{Estimation of the error and diagnostics}
\label{subsec:error}

IS with INLA can be regarded as a particular type of IS in which INLA is used to integrate most of the latent effects and hyperparameter out, so that IS is applied to the low-dimensional parameter space of $\bm{z}_{c}$. As a result, IS weights are based on the conditional (on $\bm{z}_{c}$) marginal likelihood, which is estimated with INLA. 

Similarly to what \cite{GomezRubioRue:2018} point out for the case of INLA within MCMC, it may be difficult to provide an accurate estimate of the estimation error of IS with INLA. Instead, we will argue that the estimates of the marginal likelihood provided by INLA are accurate, as several authors have discovered in a wide range of applications. In particular, \cite{HubinStorvik:2016b} have conducted a thorough analysis and they have found the estimates to be very accurate.  See, for example, \cite{GomezRubioPalmiPerales:2019,GomezRubioetal:2020} and the references therein for other uses of the marginal likelihood estimated with INLA to fit different types of spatial models with success.

Hence, we may argue that the conditional marginal likelihoods are estimated with a tiny error, and that this leads to the error introduced when computing importance weights to be small as well. Furthermore, as weights are computed by averaging over a large number of values and then re-scaling, any error introduced is likely to fade out. This should make inference on $\bm{z}_{c}$ accurate and reliable.

The error when estimating the posterior marginals of the elements in $\bm{z}_{-c}$ is also difficult to estimate as this is obtained by using a convex combination of some posterior marginals obtained by conditioning on $\bm{z}_{c}$. Again, we do not expect the error to be large as the conditional marginals are usually estimated with a very small error by INLA, and the weights are likely to have a tiny error, as discussed above.

The first example in Section~\ref{sec:examples} has been specifically conducted to assess how accurate IS-INLA is when estimating the different posterior marginals of the parameters in the model. As it can be seen, the results provide compelling evidence as to the accuracy of the estimates for the posterior marginals of the elements of $\bm{z}_c$ and $\bm{z}_{-c}$.

However, it is clear that the number of samples used in IS-INLA is crucial. For this reason, a number of numerical and graphical criteria should be used to assess that there is sufficient sample size as to provide accurate estimates. \cite{Owen:2013} describes different ways to compute the effective sample size using the importance weights, as we have stated at the end of  Section~\ref{sec:IS}. 
\cite{Elviraetal:2018} also discuss the estimation of an effective sample size for IS and make a number of important statements about how to compute this. First of all, the effective sample size must be computed separately for each function $h(x)$ involved in IS, i.e., the sample size cannot only be computed based on the weights. 

Most importantly, they state that the probability distribution $\pi(x)$ (i.e., the target distribution) is approximated by a random measure based on the sampled values of $x$ and their associated weights. Hence, the discrepancy between the sampling distribution $g(x)$ and $|h(x)| \pi(x)$ is directly related to the quality of the IS estimators, with $|h(x)|$ the absolute value of $h(x)$. Hence, this discrepancy should be assessed in some way as well. Note that this evaluations can be done for each element in $\bm{z}_c$ separately.

Similarly, \cite{Owen:2013} discusses different IS diagnostics that can be used to assess that a sufficiently large sample large has been achieved and states that sample size estimation must include the $h(x)$ function. He proposes an effective sample size criterion dependent on $h(x)$ based on the following weights:

$$
\tilde{w}_i(h) = \frac{|h(x_i)| \pi(x_i)/g(x_i)}
{\sum_{i=1}^n |h(x_i)| p(x_i)/g(x_i)}
$$

The effective sample size, dependent on $h(x)$, is

$$
n_e(h) = \frac{1}{\sum_{i=1}^n \tilde{w}_i(h)^2}
$$
\noindent
This can be computed for each of the elements in $\bm{z}_c$ so that a different per-variable effective sample size is obtained. In this particular case, $h(x)$ is taken as the identity function.

As stated above,  \cite{Elviraetal:2018} note that the IS sample and weights are implicitly used to estimate the joint posterior distribution of
$\bm{z}_c$ and their respective posterior marginals. The estimation of these posterior marginals can be regarded as the estimation of the quantiles of the posterior marginal distributions, which may be difficult. For this reason, we propose a graphical assessment based on a probability plot. This is produced by computing  the empirical cumulative probability function for each element of $\bm{z}_c$ and comparing it to its theoretical value, i.e., the cumulative probability function of a discrete uniform
distribution between 1 and $n$, with $n$ the total number of samples. Departures from the identity line will indicate that the posterior marginals are not correctly estimated.

The empirical cumulative distribution for $k$-th element in $\bm{z}_c$ is obtained ordering in increasing order the simulated values, and their associated weights in the same order. Then the empirical cumulative distribution is simply the cumulative sum of the re-ordered weights. These values can be compared with the corresponding values of the theoretical cumulative distribution. For example, the cumulative sum
of the reordered weights up to the $l$-th value must be compared to value $l / n$.

\section{Adaptive Multiple Importance Sampling with INLA}
\label{sec:amis}

The non-adaptive nature of the IS algorithm makes the performance of IS based inference heavily dependent of the a good choice of the sampling distribution. In Section~\ref{subsec:sampdist} we suggest  one preliminary sample step that could help locate  the proposal close to the target distribution. In practice, such step might require several trial-and-error rounds before reaching a satisfactory proposal $g_1(\cdot)$. Moreover, the $N_0$ preliminary samples are discarded, which might require significant computational costs. It would be therefore desirable to consider a more efficient design both more automatic and less wasteful of potentially valuable information.

To this end, we propose combining INLA with the adaptive multiple IS algorithm (AMIS) proposed in \cite{corneut_adaptive_2012}. This is one of several version of adaptive IS algorithms proposed in the literature \citep[see, for example,][and the references therein]{Bugalloetal:2017} that has the advantage  to employ a mixture of all past sampling distribution in the calculation of the importance weights such that samples can be kept after an adaptation. The proposal is updated several times in an automated way, in order to decrease the dissimilarity between target and proposal. 

The algorithm starts with a proposal distribution $g_0(\cdot)$ (here we will use Gaussian or Student-$t$) with parameters $\bm{\lambda}_0 = (\bm{\mu}_0,\bm{\Sigma}_0)$. At each iteration $t = 0,1,\dots,T$, $N_t$ samples are produced and a new, updated proposal $g_t(\cdot)$ with parameters  $\bm{\lambda}_t = (\bm{\mu}_t,\bm{\Sigma}_t)$ is computed. The new parameters are computed similarly to what is done in Section~\ref{subsec:sampdist} by matching the estimated moments of the target. 

At each step, the proposal distribution $\psi_t(\bm{z}_c)$ is then a mixture:
\begin{equation}
    \psi_t(\bm{z}_c)  =  
    \frac{\sum_{i=0}^t N_i g_{\bm{\lambda}_i}(\cdot)}{\sum_{i=1}^t N_i} = \sum_{i=0}^t \rho_i g_{\bm{\lambda}_i}(\cdot).
    \label{eq:amismix1}
\end{equation}
where $\rho_i = N_i/\sum_{i = 1}^t N_i $ is the fraction of samples generated in iteration $i$. Let  $\bm{z}_c^{(i,j)}\sim g_i(\cdot)$ be the $j$th sample generated in the  $i$th iteration; then, the corresponding importance weight is
\begin{equation}
    w^{(i,j)} = \frac{1}{N}\frac{\pi(\bm{z}_c^{(i,j)}|\mathbf{y})}{\psi_t( \bm{z}_c^{(i,j)})} 
    \propto \frac{\tilde{\pi}(\mathbf{y}|\bm{z}_c^{(i,j)})\pi(\bm{z}_c^{(i,j)})}{\psi_t( \bm{z}_c^{(i,j)})},
\end{equation}
where $\tilde{\pi}(\mathbf{y}|\bm{z}_c^{(i,j)})$ is the conditional marginal likelihood approximated with R-INLA and $\pi(\bm{z}_c^{(i,j)})$ the prior for $\bm{z}_c$ evaluated at $\bm{z}_c^{(i,j)}$.

Note that the mixture changes after every adaptation and, thereby, the weighing must be updated for all prior samples before estimating new moments for the sampling distribution. To avoid unnecessary calculations a helper variable of the numerator in Equation~\eqref{eq:amismix1} is used in the implementation. The full algorithm is shown in Algorithm~\ref{alg:amis}.

\begin{algorithm}
\vspace{5pt}
    - Initialize $N_0,N_1,\dots,N_T$, $g_{\bm{\lambda}_0}(\cdot)$, $\pi(\bm{z}_c)$\\
    \For{$t$ from $0$ to $T$}{
        \For{$j$ from $1$ to $N_t$}{
            - Generate sample $\bm{z}_c^{(t,j)} \sim g_{\bm{\lambda}_t}(\cdot)$\\
            - Fit  INLA to the model conditional on $\bm{z}_c =\bm{z}_c^{(t,j)}$. This produces the quantities: \vspace{-10pt}
            $$\tilde{\pi}(\bm{y}|\bm{z}_c^{(t,j)})\textrm{   and   }\tilde{\pi}(z_{-c,i}|\bm{y},\bm{z}_c^{(t,j)}), \forall z_{-c,i}\in\bm{z}_{-c}$$
            - Compute:\vspace{-13pt}
            $$\gamma^{(t,j)} = \sum\limits_{l=0}^t N_l\cdot g_{\bm{\lambda}_t}(\bm{z}_c^{(t,j)})\hspace{5pt} \textrm{and}\hspace{5pt} w^{(t,j)}  = \frac{\tilde{\pi}(\bm{y}|\bm{z}_c^{(t,j)}) \pi(\bm{z}_c^{(t,j)})} { \left[\gamma^{(t,j)}\!\middle/\!\sum_{l=0}^t N_l\right]}$$ \\ 
        }
        
        \If{$t > 0$}{
            \For{$l$ from $0$ to $t-1$}{
                \For{$j$ from $1$ to $N_l$}{
                    - Update past importance weights:\\\vspace{-14pt}
                    $$\gamma^{(l,j)} \gets \gamma^{(l,j)} + N_t g_{\bm{\lambda}_t}(\bm{z}_c^{(l,j)}) \hspace{5pt} \textrm{and}\hspace{5pt} w^{(l,j)} \gets \frac{\tilde{\pi}(\bm{y}|\bm{z}_c^{(l,j)})\pi(\bm{z}_c^{(l,j)})}{ \left[\gamma^{(l,j)}\!\middle/\!\sum_{k=0}^t N_k\right]}$$\\
                }
        }
        }
        - Calculate $\bm{\lambda}_{t+1}$ using the weighted set of samples:\vspace{-5pt}
         $$(\{\bm{z}_c^{(0,1)},w^{(0,1)}\},\dots,\{\bm{z}_c^{(t,N_t)},w^{(t,N_t)}\})$$ \\
    }
    - Estimate $\pi(\bm{z}_c|\bm{y})$ using kernel density estimation\\
    - Estimate posterior marginals of $\bm{z}_{-c}$:
        $$
        \tilde{\pi}(z_{-c,i}\given\bm{y}) = \left.\sum\limits_{t=0}^T\sum\limits_{j=1}^{N_t} w^{(t,j)}\tilde{\pi}(\bm{z}_{-c,i}\given\bm{y},\bm{z}_c^{(t,j)})\!\middle/\!\sum\limits_{t=0}^T\sum\limits_{j=1}^{N_t} w^{(t,j)}\right.
        $$
\caption{A detailed description of the AMIS-INLA algorithm}
\label{alg:amis}
\end{algorithm}

\section{Examples}
\label{sec:examples}
In this section we present a series of examples to illustrate the methods proposed in the previous sections. The first three examples are taken from \cite{GomezRubioRue:2018}. If not stated otherwise, the same strategy for running  IS-INLA and AMIS-INLA will be used: they both start from the same preliminary proposal distribution, a Gaussian or Student-$t$ distribution with 3 degrees of freedom with location $\bm{\mu}_0$ and scale $\bm{\Sigma}_0$. IS-INLA uses then 800 samples to update the proposal and estimate the new parameters  $\bm{\mu}_1$ and $\bm{\Sigma}_1$. The preliminary 800 samples are then discarded and  10000 samples are generated from the new proposal distribution. AMIS-INLA generates a total of 10000 samples by adapting the proposal distribution $T = 27$ times, to have a high number of adaptation steps. At each adaptation step $N_t$ samples ( $t = 1,\dots,T$) are produced. $N_t$ varies between 250 and 500. No sample is discarded. For the MCMC-INLA algorithm, we collect 10000 samples after convergence has been reached.

In these experiments a computer with a total of 28 CPUs with 3.2 GHz clock speed, where a fixed number of 10 cores were used to prevent any major deviations in the computation speeds caused by the parallelization. All our implementations and experiments are publicly available in the repository (\url{https://github.com/berild/inla-mc}).


\input{toy}
\input{lasso}
\input{missing}

\input{pqr}

\section{Discussion}\label{sec:discussion}

The integrated nested Laplace approximation is a suitable approach for approximate Bayesian inference for latent Gaussian models, as described in \cite{isi:000264374200002}. Extending the use of INLA to other classes of models has been considered by several authors using INLA together with numerical integration or MCMC methods. Here we have illustrated a novel approach to extend the models that INLA can fit by combining importance sampling and adaptive multiple importance sampling with INLA.

This new approach has a number of advantages over other similar approaches. First of all, importance sampling is a very simple algorithm that can also be easily parallelized, leading to a huge computational speed up. This means that, in practice, times for model fitting remain small. In the examples developed in this paper we have illustrated how IS and AMIS with INLA are able to fit a wide range of models. Furthermore, the numerical experiments conducted show that the approximations of the posterior marginals obtained with IS and AMIS with INLA are also accurate and close to the actual posterior marginals.

This paper also discusses numerical and graphical diagnostics to assess the accuracy of IS/AMIS when used in combination with INLA to fit models. We have observed that the different criteria usually agree, with small effective sample sizes associated to poor estimates of the posterior marginal distribution of some the model parameters. Hence, these criteria can effectively be used to critically assess the quality of the estimates produce by IS/AMIS with with INLA. In this sense, in the examples developed in the paper AMIS seemed to provide better estimates when used in combination with INLA for model fitting.

%


\bibliographystyle{chicago}
\bibliography{local}

\end{document}

%% file: martin-math-commands.tex
\newcommand{\reals}{\mathbb{R}} 
\newcommand{\dd}{\mathrm{d}}

\newcommand{\given}[1][]{\:#1\vert\:} 


%% file: toy.tex
\subsection{Bivariate linear model}
\label{subsec:toy}

In the first example, we repeat the simulated study in \cite{GomezRubioRue:2018} and consider a  simple linear model. %
$100$ responses are simulated from
\begin{equation*}
    y_i = \beta_0 + \beta_1 x_{1i} + \beta_2 x_{2i} + \epsilon_i,\textrm{  for } i = 1,\dots,100.
\end{equation*}
Covariates $\bm{x}_1$ and $\bm{x}_2$ are simulated from a uniform distribution between 0 and 1 while the error terms $\epsilon_i$ are simulated from a standard normal distribution (i.e., precision is $\tau = 1$). Moreover, we set  $\beta_0 = \beta_1 = 1$,  and $\beta_2 = -1$.

This model can be easily fitted using INLA, and since the likelihood is Gaussian, results are exact up to an integration error. \cite{GomezRubioRue:2018} use this example to compare the MCMC-INLA approximations with the exact INLA results and to show how MCMC-INLA gives also access to some joint posterior inference, for example the joint posterior of $\beta_1$ and $\beta_2$ that INLA cannot provide. We repeat this example to show that both IS-INLA and AMIS-INLA can reach the same results in just a fraction of the time used by MCMC-INLA.

\begin{figure}[!ht]
    \centering
    \includegraphics[width=0.99\textwidth]{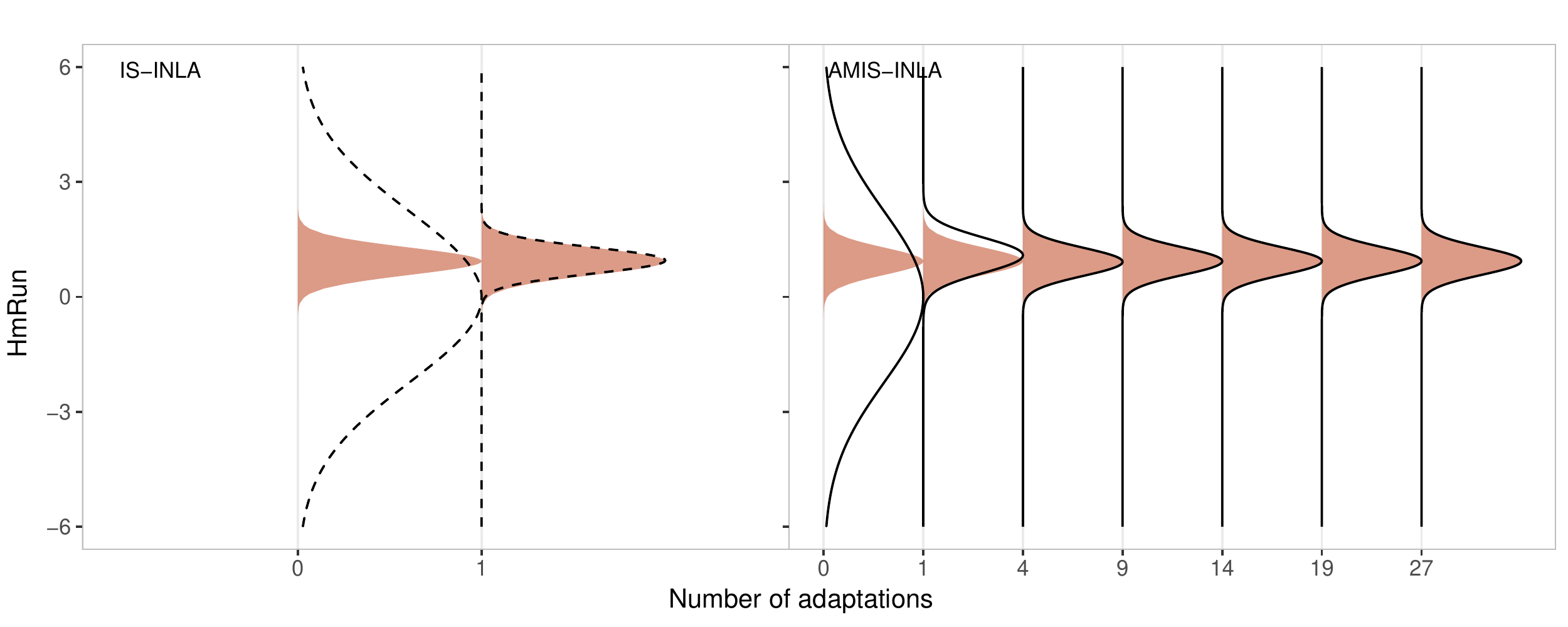}
    \caption{A visual representation of the initial search in IS-INLA (\dashed, left) and the adaptation of proposal distribution in AMIS-INLA (\solid, right) for $\beta_1$ in the bivariate linear model. The $x$-axis is the number of adaptations of the proposal distribution. The lines (\solid, \dashed) are the proposal distributions and the filled area (\square) denotes the target density.}
        \label{fig:toy_adapt}
\end{figure}

For this model we have  $\bm{z} = (\beta_0,\beta_1,\beta_2,\tau)$, and we set $\bm{z}_c=(\beta_1,\beta_2)$ and $\bm{z}_{-c}=(\beta_0,\tau)$.
As in \cite{GomezRubioRue:2018}, the proposal in MCMC-INLA is a bivariate Gaussian with mean   equal to the previous state $\bm{\beta}^{(j)}$ and variance of $0.75^2\cdot\mathbf{I}$. We set   $\bm{\beta}^{(0)} = \bm{0}$ as starting value.
Both IS-INLA and AMIS-INLA use as first proposal distribution a bivariate Gaussian with mean  $\bm{\mu}_0 = \bm{0}$ 
and covariance  $\bm{\Sigma}_{0} = 5 \cdot \mathbf{I}$.  Figure~\ref{fig:toy_adapt} (a-b)
show how the initial proposal distribution for $\beta_1$ changes after the preliminary step in IS-INLA and during the adaptation process in AMIS-INLA. In this case the preliminary step in IS-INLA seems to be sufficient to correctly locate the target. The adaptation process in AMIS-INLA could have been stopped earlier giving faster computing time.

\begin{figure}[!ht]
     \centering
     \begin{subfigure}[a]{0.49\textwidth}
         \centering
         \includegraphics[trim=6pt 9pt 5pt 5pt, clip,width=\textwidth]{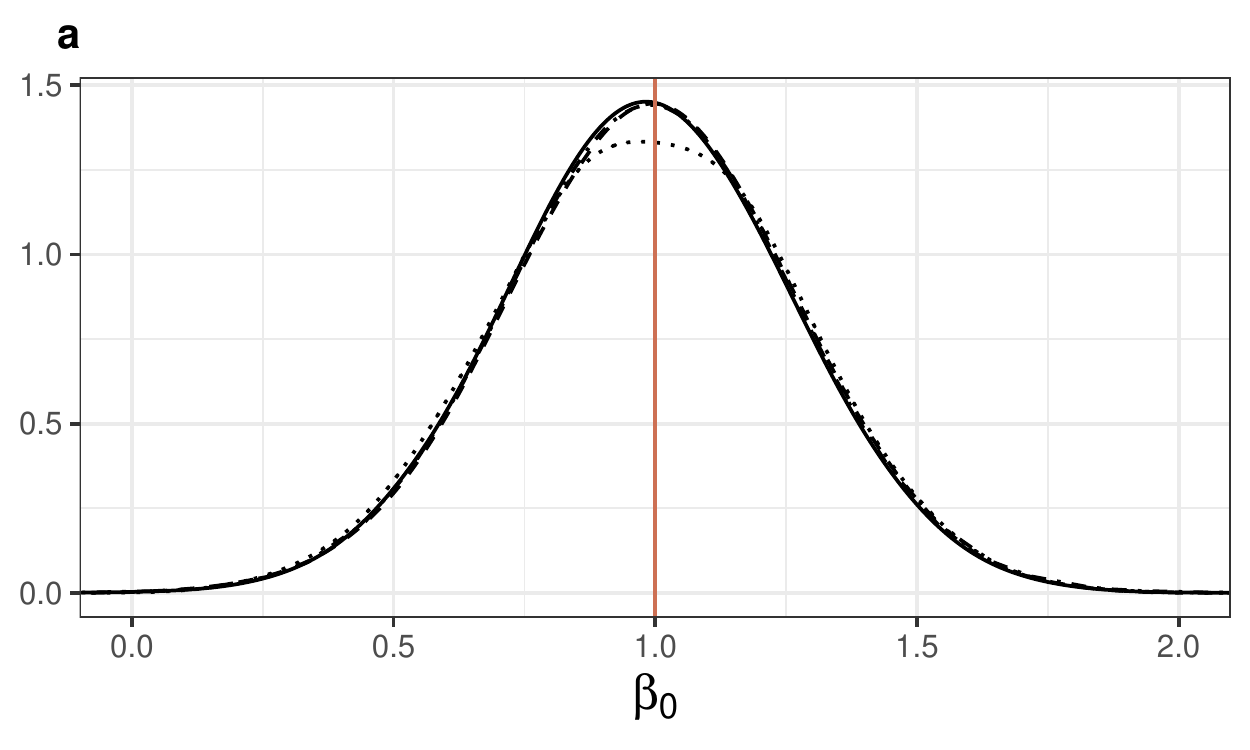}
     \end{subfigure}
     \begin{subfigure}[a]{0.49\textwidth}
         \centering
         \includegraphics[trim=6pt 9pt 5pt 5pt, clip,width=\textwidth]{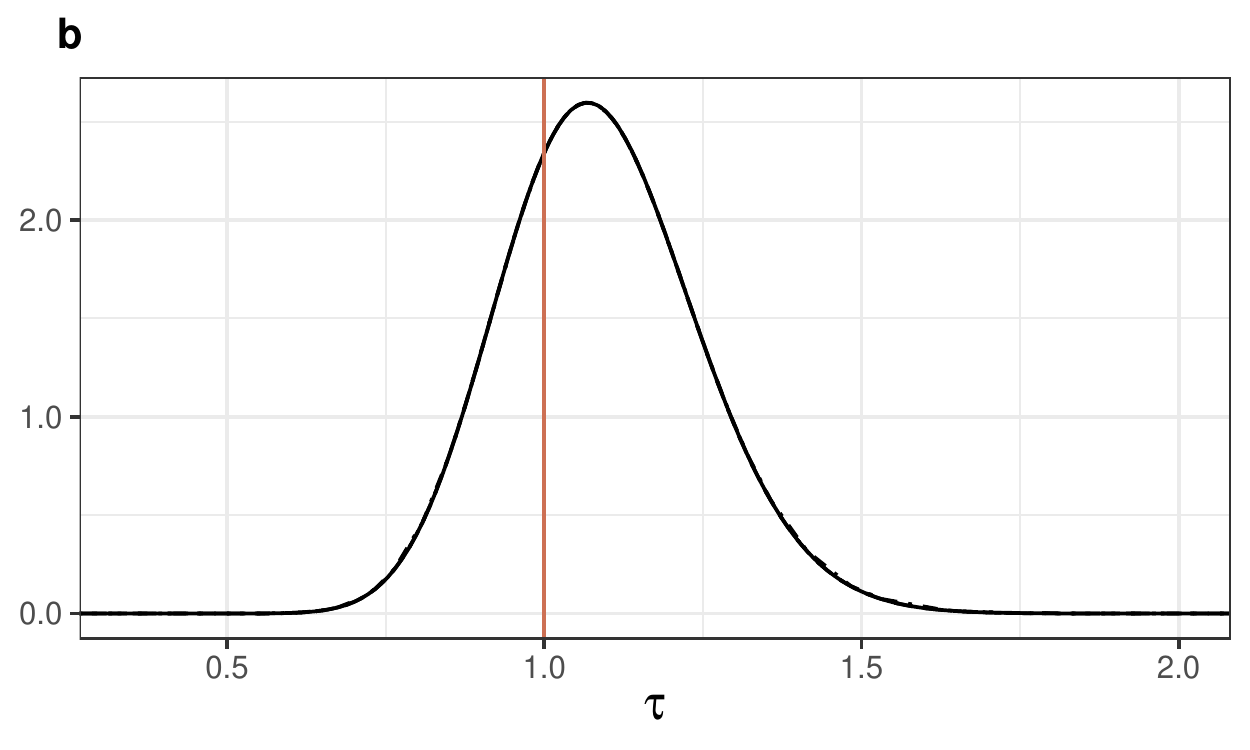}
     \end{subfigure}
     \begin{subfigure}[a]{0.49\textwidth}
         \centering
         \includegraphics[trim=6pt 9pt 5pt 5pt, clip,width=\textwidth]{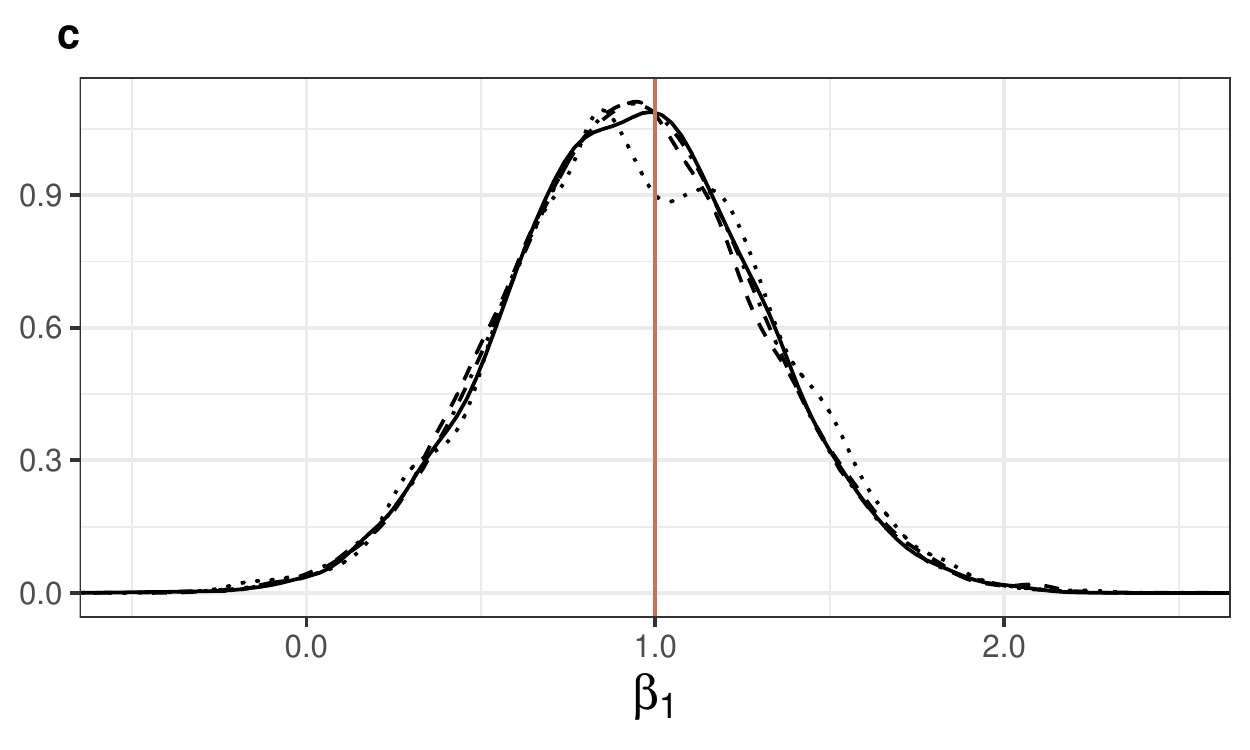}
     \end{subfigure}
     \begin{subfigure}[a]{0.49\textwidth}
         \centering
         \includegraphics[trim=6pt 9pt 5pt 5pt, clip,width=\textwidth]{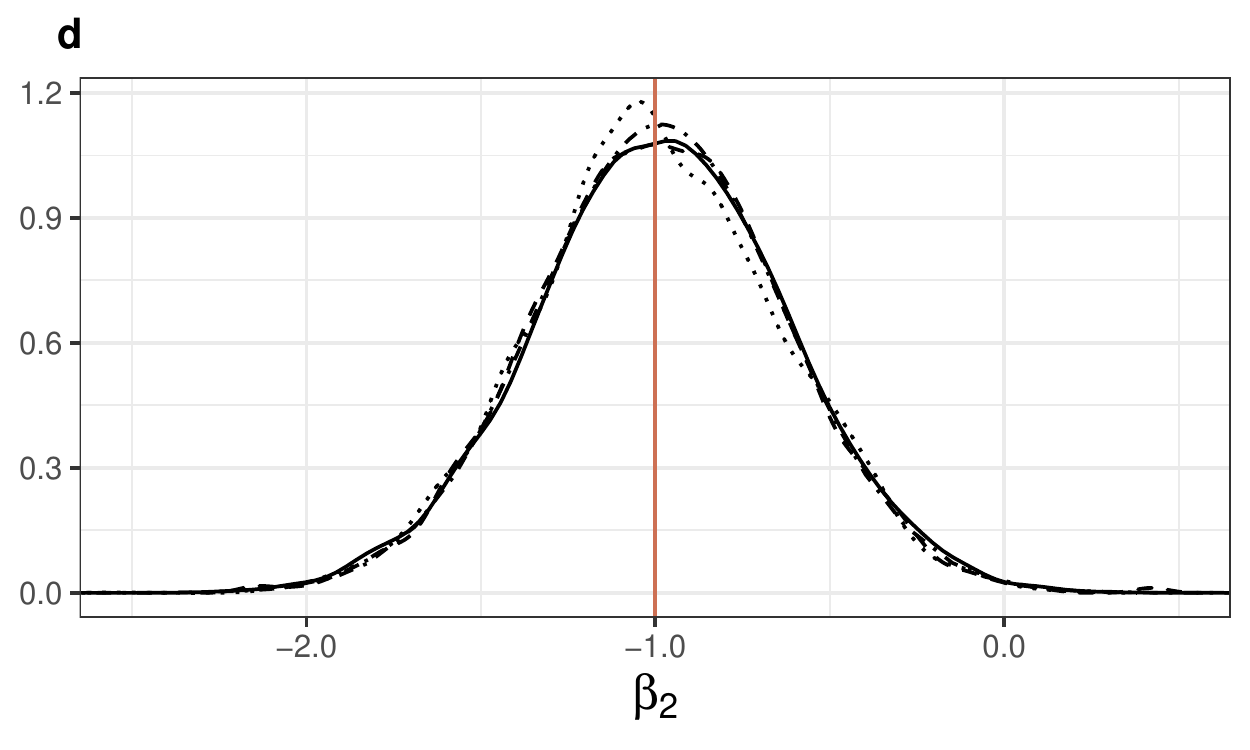}
     \end{subfigure}
        \caption{Posterior marginals of all parameters in the bivariate linear model approximated with AMIS-INLA (\solid), IS-INLA (\dashed), MCMC-INLA (\dotted), and INLA (\dotdash). The line (\vsolid) is the value of the parameter chosen for the simulation of data.}
        \label{fig:toy_uni}
\end{figure}

\begin{figure}
     \centering
     \begin{subfigure}[b]{0.49\textwidth}
         \centering
         \includegraphics[trim=3pt 9pt 5pt 5pt, clip,width=\textwidth]{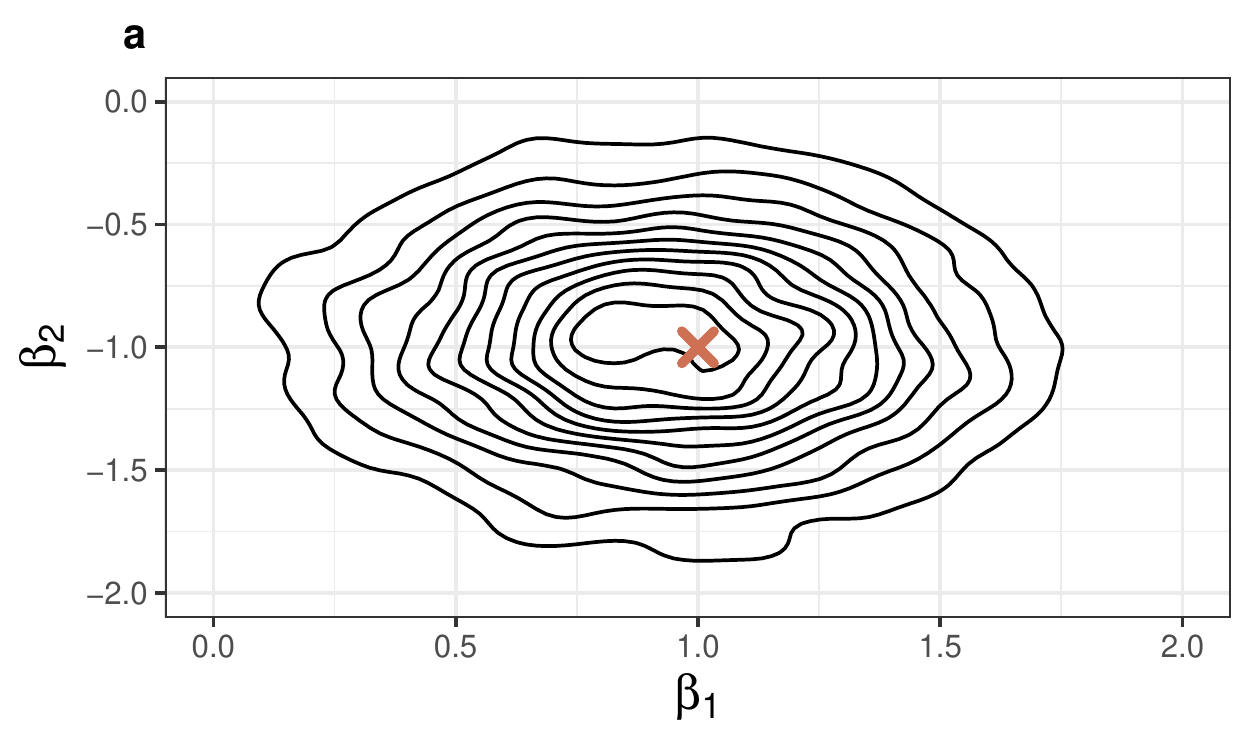}
     \end{subfigure}
     \begin{subfigure}[b]{0.49\textwidth}
         \centering
         \includegraphics[trim=3pt 9pt 5pt 5pt, clip,width=\textwidth]{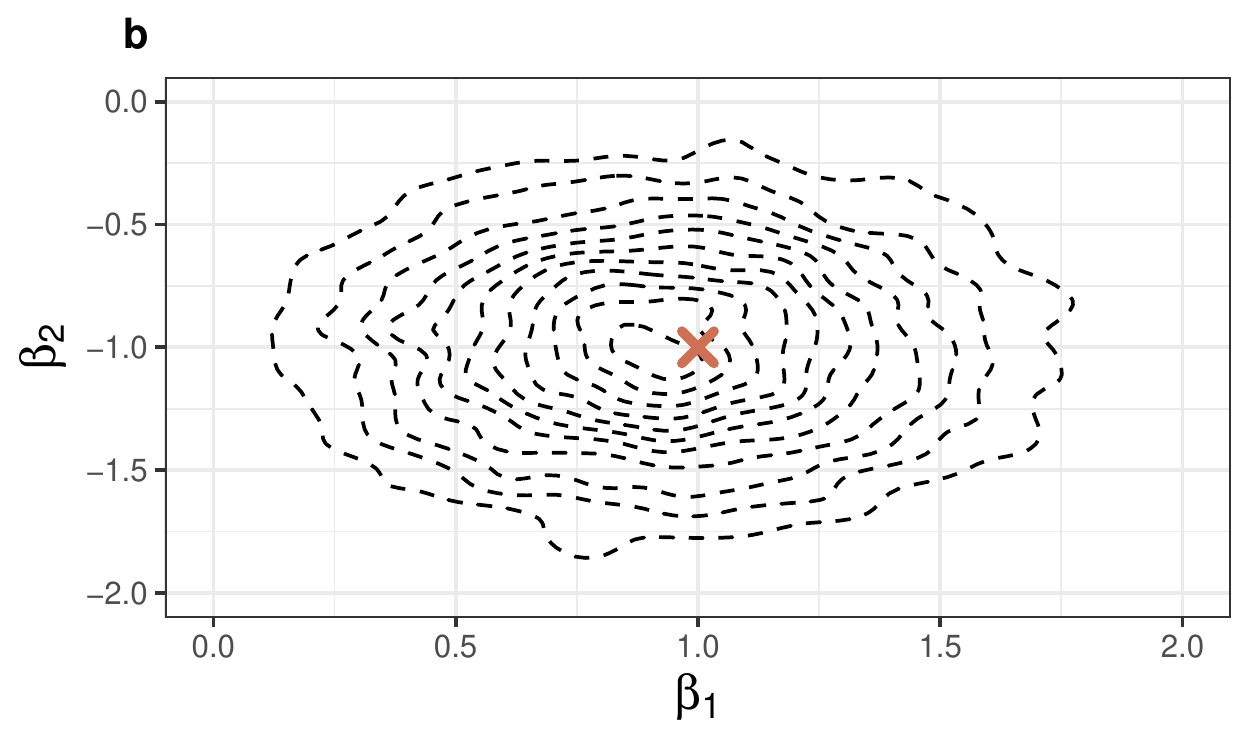}
     \end{subfigure}
     \begin{subfigure}[b]{0.49\textwidth}
         \centering
         \includegraphics[trim=3pt 9pt 5pt 5pt, clip,width=\textwidth]{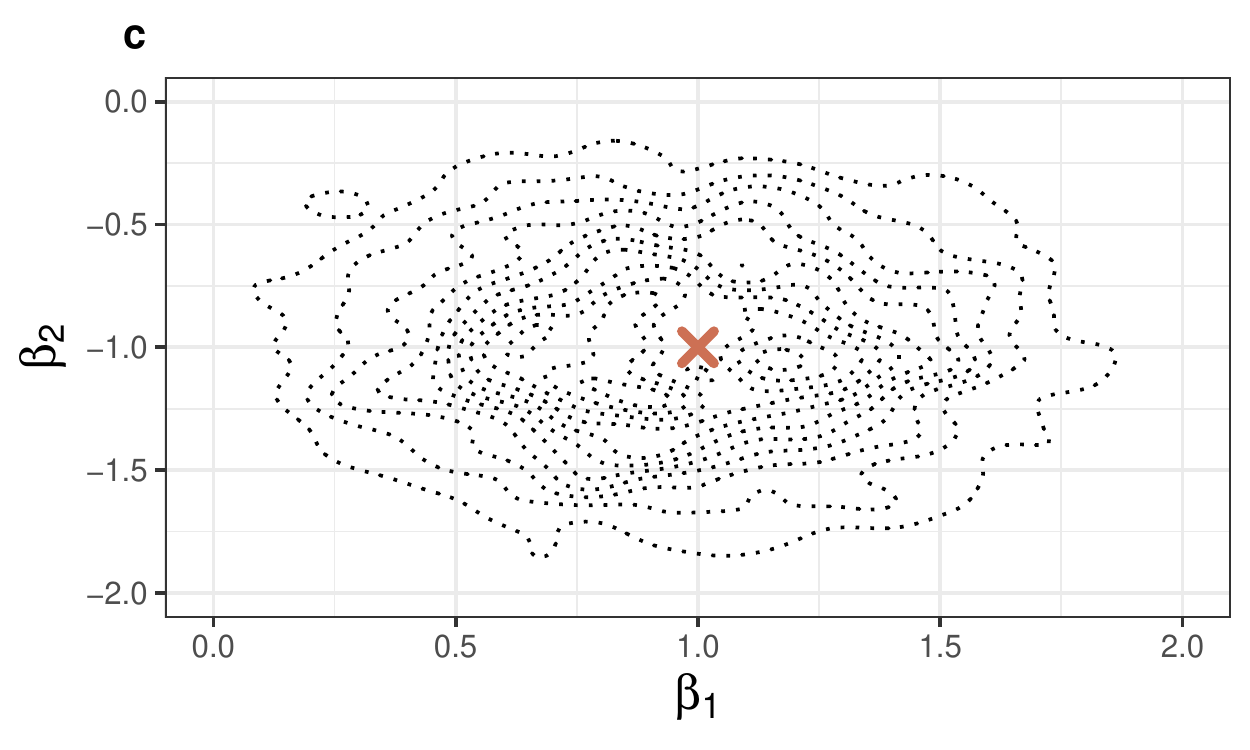}
     \end{subfigure}
     \begin{subfigure}[b]{0.49\textwidth}
         \centering
         \includegraphics[trim=3pt 9pt 5pt 5pt, clip,width=\textwidth]{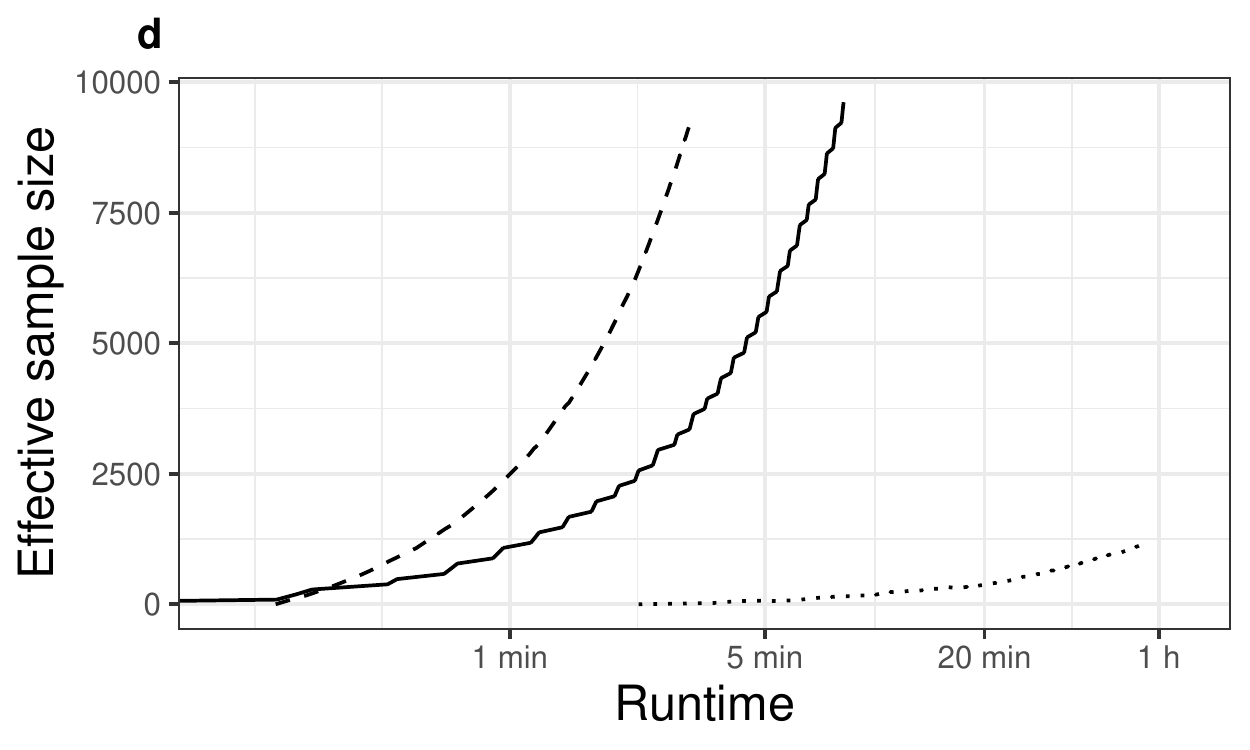}
     \end{subfigure}
    \caption{The joint posterior distribution of $\bm{\beta}$ in the bivariate linear model obtained using AMIS-INLA (\solid), IS-INLA (\dashed), MCMC-INLA (\dotted), and bottom right the running effective sample size of the respective methods. The (\cross) denotes the values of $\bm{\beta}$ chosen for the simulation of data.}
        \label{fig:toy_joint}
\end{figure}

Figure~\ref{fig:toy_uni} shows  the approximated posterior marginals of $\beta_0$, $\beta_1$, $\beta_2$, and $\tau$ from the combined approaches, while Figure~\ref{fig:toy_joint} (a-c) show the estimated joint posterior for $(\beta_1,\beta_2)$. Posterior marginals from INLA alone and  true values of the parameters are included for reference. All methods seem to be able to recover the parameters. MCMC-INLA seems to be the method most affected by Monte Carlo error,  visible both in marginals and  joint distributions.

Figure~\ref{fig:toy_joint} (d) shows the running ESS, as in Equation \eqref{eq:ess} for all combined approaches. Clearly, MCMC-INLA  has achieved  fewer effective samples in longer time.  IS-INLA, which in this case is the most efficient method,   achieved 49.2 effective samples per second, AMIS-INLA  19.5 effective samples per second,  MCMC-INLA managed only  0.35 effective samples per second. 

Finally, the different numerical and graphical diagnostics discussed in Section~\ref{subsec:error} have been computed to assess the quality of the estimates provided by IS with INLA. The per-variable sample sizes $n_e(h)$ for $\beta_1$ and $\beta_2$ are 8124.978 and 7555.027 for IS-INLA and 8510.857 and 8476.343 for AMIS-INLA. Similarly, the probability plots provide a curve that is very close to the identity line, which points to a very good estimate of the posterior marginal distributions. These are not shown here but provided in the Supplementary Materials.

%% file: lasso.tex
\subsection{Bayesian Lasso}
\label{subsec:lasso}

The Lasso is a popular linear regression method that also provides variable selection \citep{tibshirani_regression_1996}. 
For a  model with Gaussian likelihood, the Lasso tries to estimate the regression  coefficients by minimizing
\begin{equation}
    \sum_{i=1}^N \left(y_i - \alpha - \sum_{j=1}^{n_\beta} \beta_j x_{ji}\right)^2 + \lambda\sum_{j=1}^{n_\beta}|\beta_j|,
    \label{eq:lassoopt}
\end{equation}
where $y_i$ is the response variable,  and $x_{ji}$ the associate covariates. $N$ is the number observations and $n_\beta$ the number of covariates. %
The shrinkage of the coefficients is controlled by the regularization parameter $\lambda>0$. Larger values of $\lambda$ results in larger shrinkage i.e. coefficients tend more towards zero. Using $\lambda = 0$ would yield the maximum likelihood estimates. 

In a Bayesian setting, the Lasso can be regarded as a standard regression model with Laplace priors on the variable coefficients. The  Laplace distribution is 
\begin{equation*}
    f(\beta) = \frac{1}{2\sigma}\exp\left(-\frac{|\beta-\mu|}{\sigma}\right).
\end{equation*}
where $\mu$ is a location parameter and $\sigma>0$ a scale parameter corresponding to the inverse of the regularization parameter $\sigma = 1/\lambda$. The Laplace prior is not available for the latent field in R-INLA, but the model is simple to fit if we condition on the values of the $\bm{\beta}$ coefficients.

We use the {\tt Hitters} dataset \citep{james_introduction_2013}, available in the \texttt{ISLR} R package \citep{ISLR},
 that contains several statistics about players in the Major League Baseball, including salary in 1987.  
Following \cite{GomezRubioRue:2018}, we want to predict the player's salary in 1987 based on 5 variables, see  \cite{GomezRubioRue:2018} for details on the model and the choice of priors.

 MCMC-INLA uses a multivariate Gaussian proposal distribution for $\bm{\beta}^{(j)}$ with mean equal to the previous sample $\bm{\beta}^{(j-1)}$ and precision $4\cdot\mathbf{X}^T\mathbf{X}$, as \cite{GomezRubioRue:2018} reported good acceptance rates using this proposal. %
Here, $\mathbf{X}$ is the model matrix with the individual observations as rows and the different covariates as columns. We set the initial state to $\bm{\beta}^{(0)}=\bm{0}$. %
For the IS-INLA and AMIS-INLA methods, we  use a  multivariate Student-$t$ proposal  with $\nu=3$ and initial parameters $\bm{\mu}_0 = \bm{0} $ and  $\bm{\Sigma}_0 = (\mathbf{X}^T\mathbf{X})^{-1}$.

\begin{figure}[!ht]
     \centering
     \begin{subfigure}[]{0.325\textwidth}
         \centering
         \includegraphics[trim=5pt 9pt 5pt 10pt, clip,width=\textwidth]{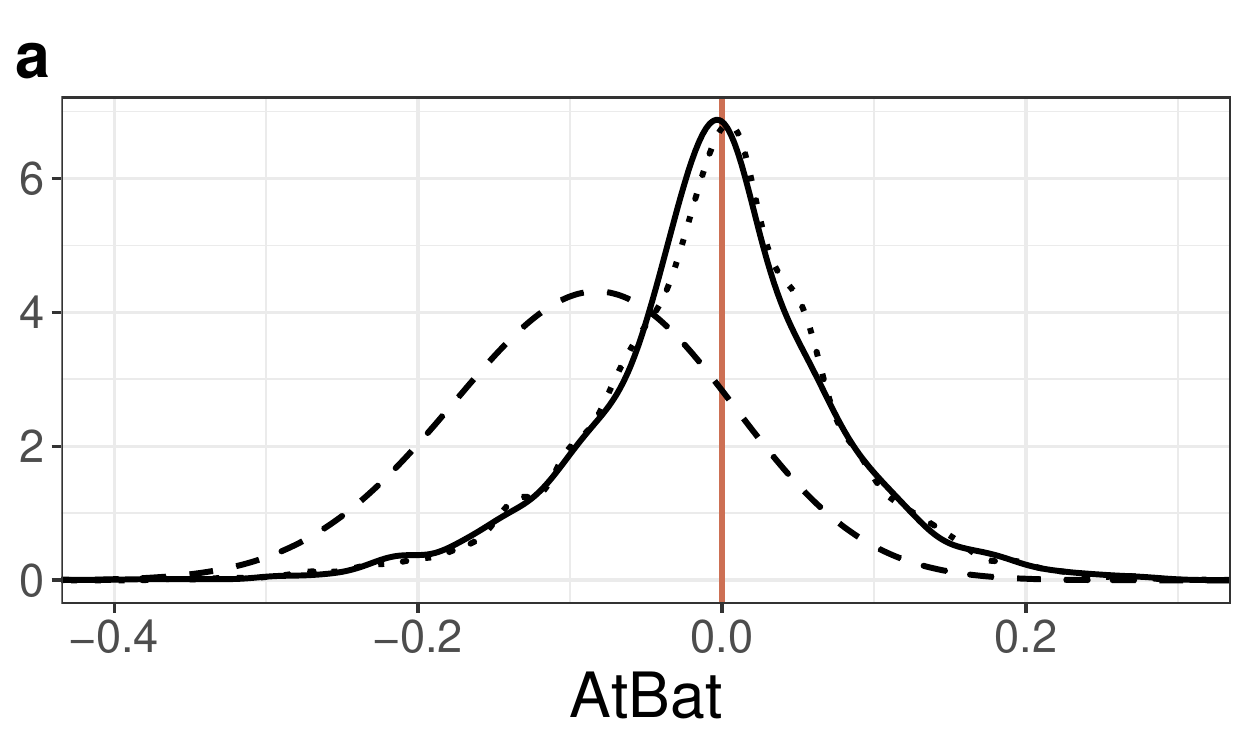}
     \end{subfigure}
     \begin{subfigure}[]{0.325\textwidth}
         \centering
         \includegraphics[trim=5pt 9pt 5pt 10pt, clip,width=\textwidth]{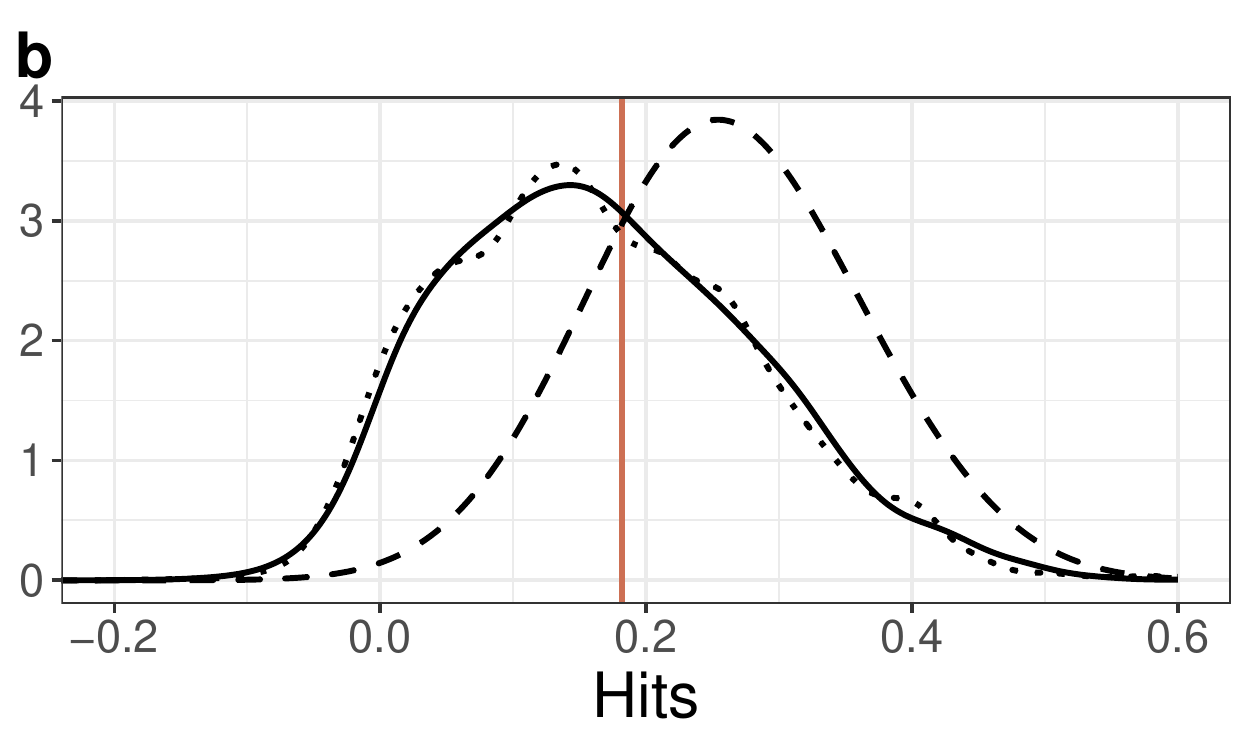}
     \end{subfigure}
     \begin{subfigure}[]{0.325\textwidth}
         \centering
         \includegraphics[trim=5pt 9pt 5pt 10pt, clip, width=\textwidth]{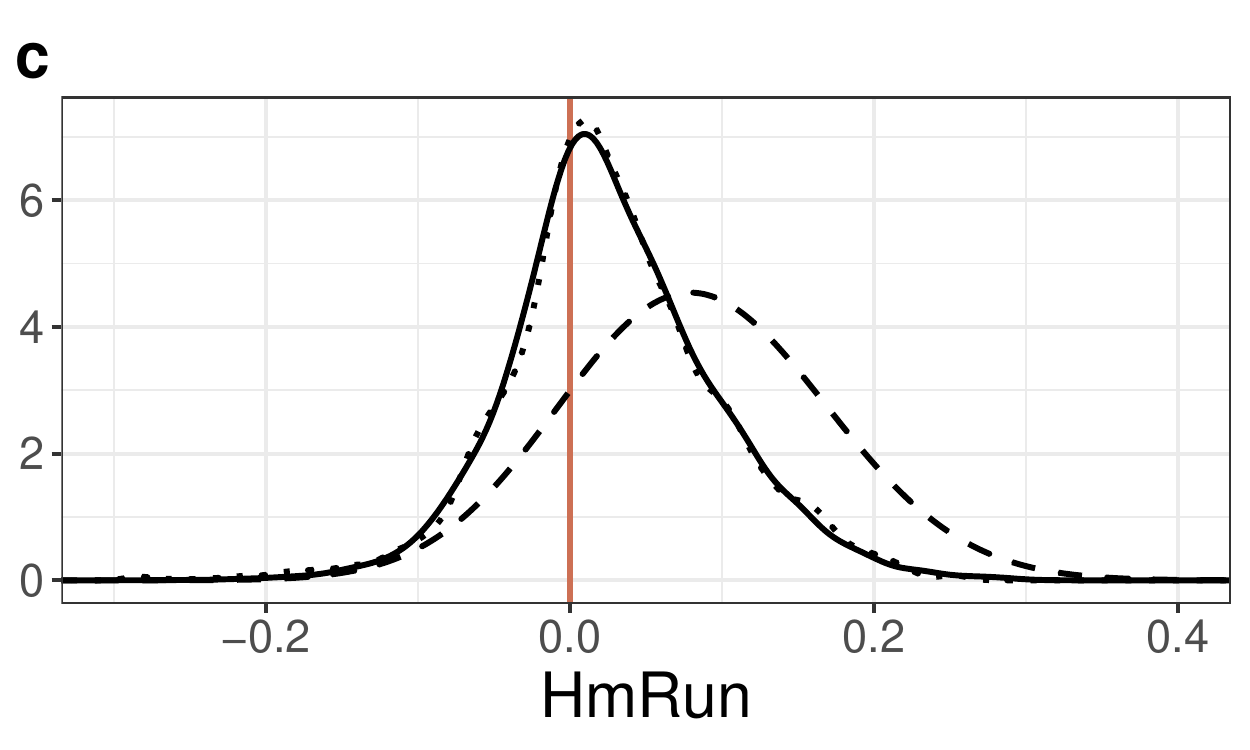}
     \end{subfigure}
      \begin{subfigure}[]{0.325\textwidth}
         \centering
         \includegraphics[trim=5pt 9pt 5pt 10pt, clip,width=\textwidth]{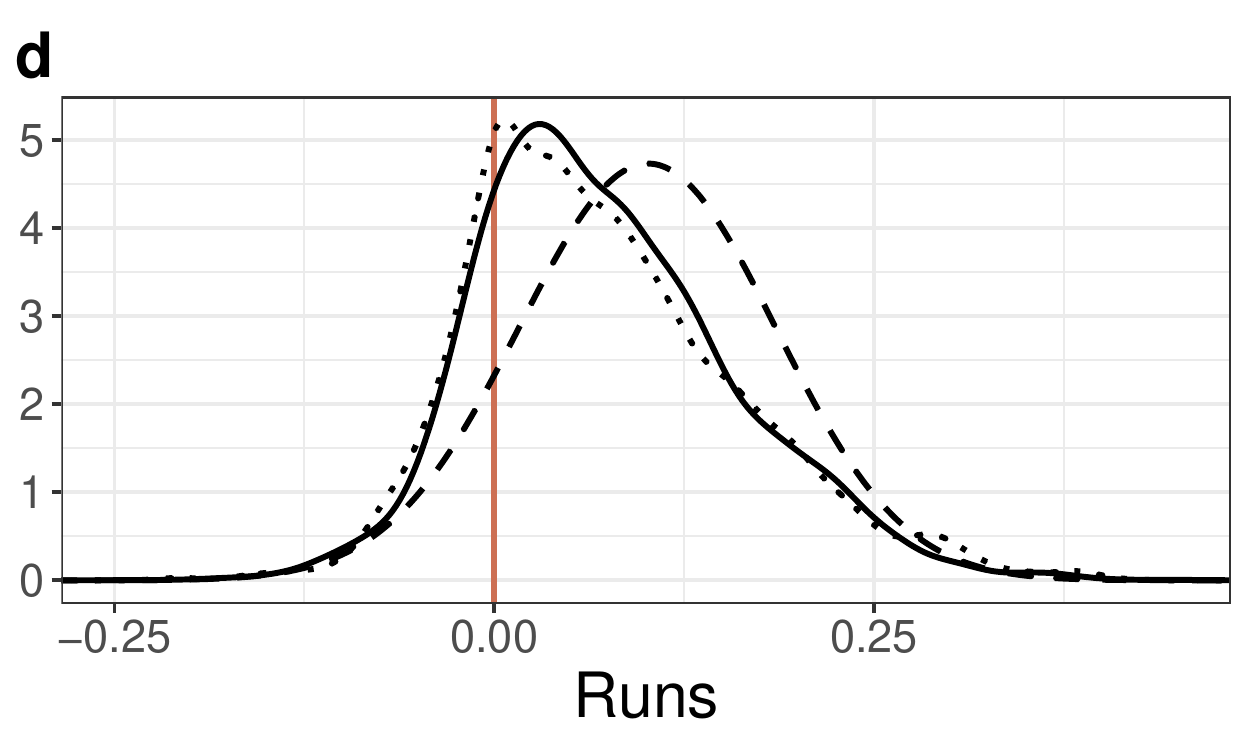}
     \end{subfigure}
          \begin{subfigure}[]{0.325\textwidth}
         \centering
         \includegraphics[trim=5pt 9pt 5pt 10pt, clip,width=\textwidth]{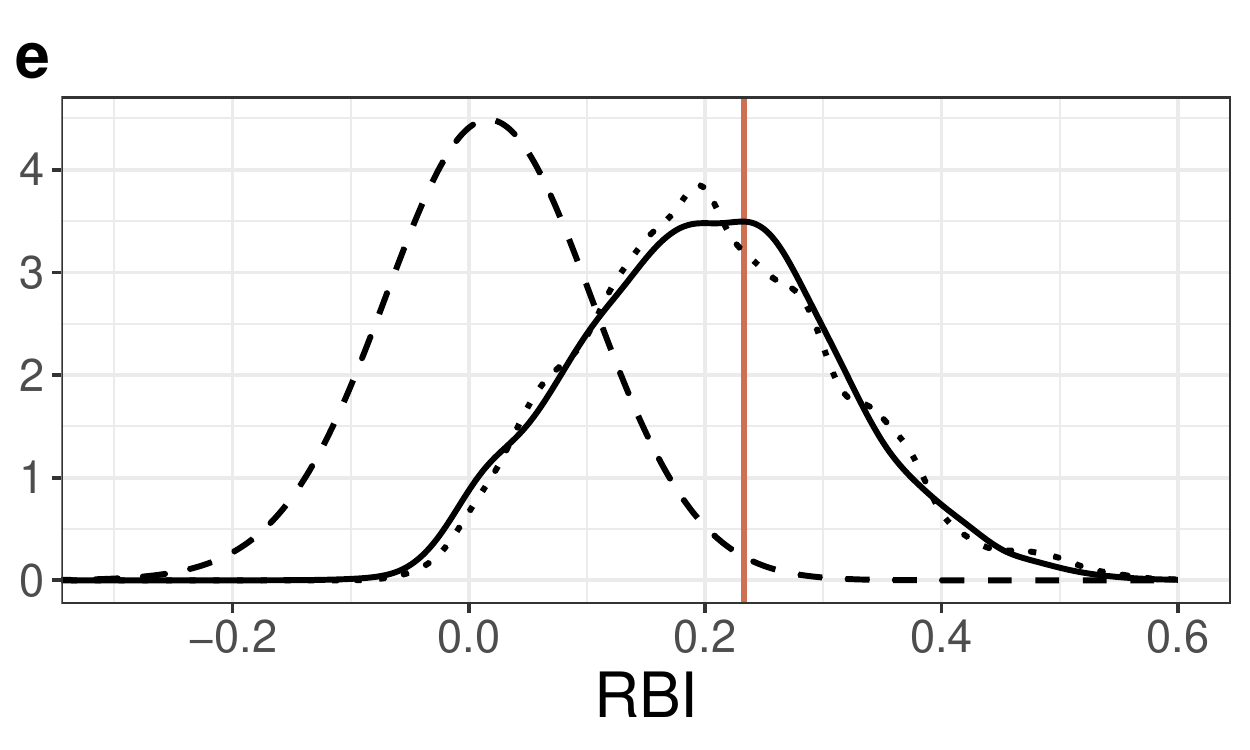}
     \end{subfigure}
     \begin{subfigure}[]{0.325\textwidth}
         \centering
         \includegraphics[trim=5pt 9pt 5pt 10pt, clip,width=\textwidth]{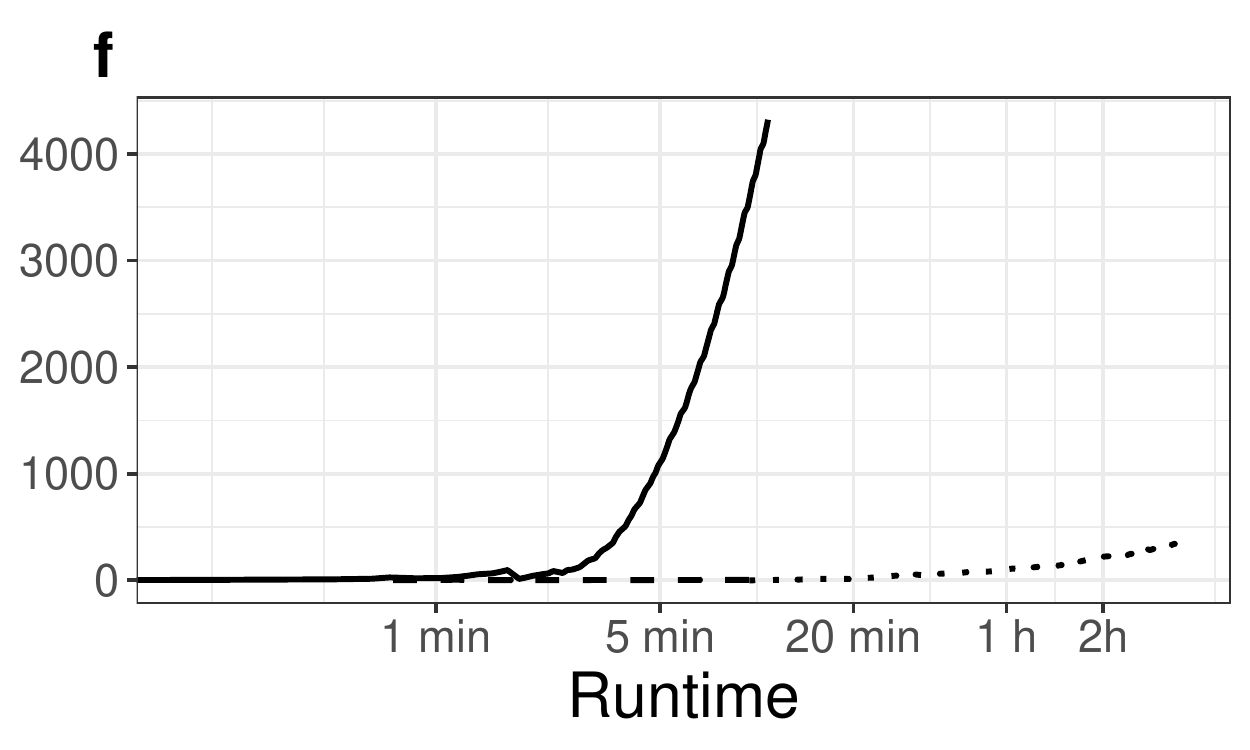}
     \end{subfigure}
    \caption{Approximate posterior marginals of the coefficients of the Bayesian Lasso model fitted with AMIS-INLA (\solid), IS-INLA (\dashed) and MCMC-INLA (\dotted), and the Lasso estimates of the coefficients (\vsolid). The running effective samples size (bottom right) where runtimes is presented in logarithmic scale.}
    \label{fig:lasso}
\end{figure}

\begin{figure}[!ht]
    \centering
    \includegraphics[width=0.99\textwidth]{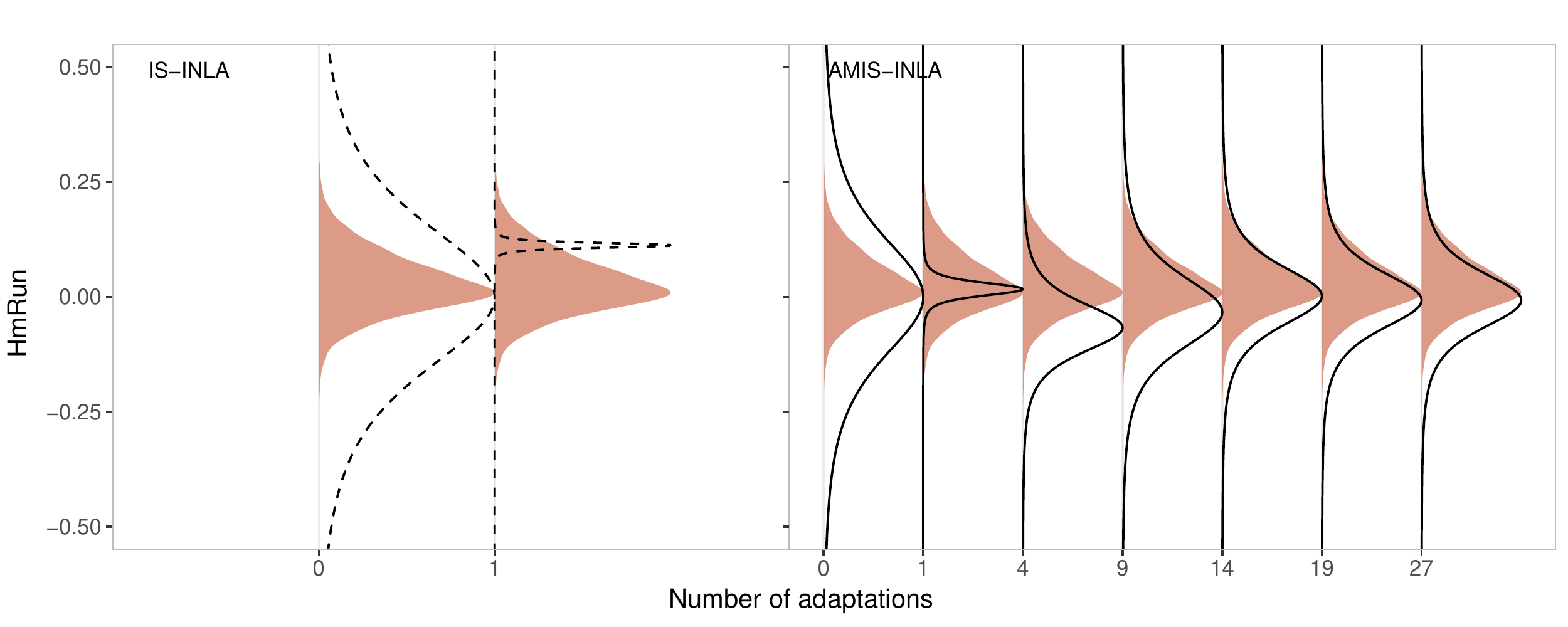}
    \caption{A visual representation of the initial search in IS-INLA (\dashed) and adaptation of the proposal distribution in AMIS-INLA (\solid) for $\mathrm{HmRun}$ in the Bayesian lasso model. The $x$-axis is the number of adaptations of the proposal distribution and the fill (\square) is the target density.}
    \label{fig:lasso_adapt}
\end{figure}

Figure~\ref{fig:lasso} (a-e) shows the estimated posterior marginals for the 5  coefficients.   MCMC-INLA and AMIS-INLA  provide similar estimates of the coefficients, with the posterior mode closely matching the Lasso regression estimates. 
On the contrary,  IS-INLA  does not provide accurate results. The problem here is that the preliminary 800 samples are not enough to correctly locate the proposal density. Figure~\ref{fig:lasso_adapt} illustrates the problem occurring when the dimensionality of $\bm{z}_c$ is high, as few good samples are obtained in the preliminary steps the variance of the estimator for the mean and variance in Equation~\eqref{eq:iscov} is large and, thus, the estimated proposal distribution is poor. We could have used more samples in the preliminary step and make the IS-INLA work, but our point here is to show that AMIS-INLA requires less tuning in order to work well. 

Figure~\ref{fig:lasso} (f) shows  the running effective sample size. We get  $\mathrm{ESS}_{\mathrm{MCMC}} = 2784$ and $\mathrm{ESS}_{\mathrm{AMIS}} = 4321$ based on their 10,000 generated samples. %
 MCMC-INLA  used 14 hours and 19 minutes to completed, whereas  AMIS-INLA  used 10 minutes and 49 seconds, resulting in an effective samples per seconds of $0.05$ and $6.65$.

\begin{figure}[!ht]
    \centering
    \includegraphics[width=0.99\textwidth]{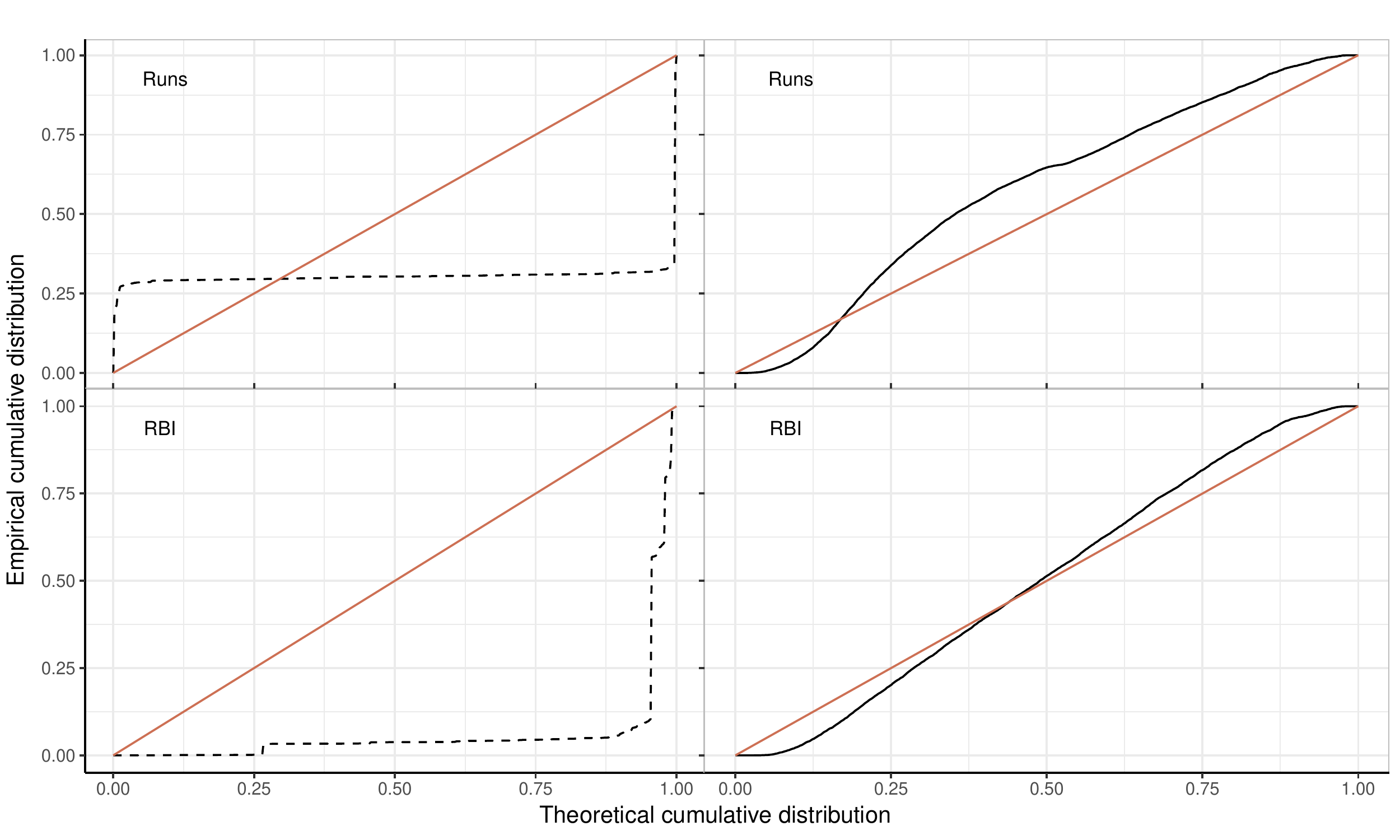}
    \caption{Probability plots for Runs and RBI parameters in the Bayesian lasso model obtained with IS-INLA (\dashed) and AMIS-INLA (\solid). The comparison line (\cdashed) denotes equivalent empirical and theoretical cumulative distributions.}
    \label{fig:lasso_pplot}
\end{figure}

Regarding per-variable diagnostics, effective sizes $n_e(h)$ for IS-INLA are about just 4 for all the coeffients, while they are between 2446.961 (for $\beta_4$) and 3243.505 (for $\beta_5$)for AMIS-INLA. This points to the fact that AMIS-INLA provides more accurate estimates in this case. Note that in this way it is possible to assess the quality of the different IS estimates. Figure~\ref{fig:lasso_pplot} shows the probability plots for $\beta_4$ and $\beta_5$ for IS-INLA and AMIS-INLA to assess the estimate of their posterior marginals
from the weights and sample. This confirms that AMIS-INLA should be preferred in this case and illustrates the use of the IS diagnostics introduced in Section~\ref{subsec:error}.

Figure~\ref{fig:lasso} (f) shows  the running effective sample size. We get  $\mathrm{ESS}_{\mathrm{MCMC}} = 2784$ and $\mathrm{ESS}_{\mathrm{AMIS}} = 4321$ based on their 10,000 generated samples. %
 MCMC-INLA  used 14 hours and 19 minutes to completed, whereas  AMIS-INLA  used 10 minutes and 49 seconds, resulting in an effective samples per seconds of $0.05$ and $6.65$.

%% file: missing.tex
\subsection{Missing Covariates}
\label{subsec:missing}

The next example is also taken from \cite{GomezRubioRue:2018} and discusses the case of imputation of missing covariates.  We consider the {\tt nhanes} dataset \citep{schafer_analysis_1997}, available in R package \texttt{mice} \citep{mice}, containing information on age, body mass index ({\tt bmi}), hypertension status ({\tt hyp}) anc cholesterol level ({\tt chl}). Cholesterol level is the response variable and there are missing values both in the response variable and in the {\tt bmi} covariate. INLA can deal with missing values in the response, but is not able to provide imputation for missing covariates.   The model set up, and the setting for the MCMC-INLA algorithm are identical to \cite{GomezRubioRue:2018} and the reader can refer to that for details.   

The initial proposal for IS-INLA and AMIS-INLA is a multivariate Gaussian $\bf{\mu}_0 = \mu_0\bf{1}$ and $\bf{\Sigma}_0 = \sigma_0\bf{I}$ where $\mu_0$ is the mean of the observed covariates and $\sigma_0$ is twice the standard deviation of the observed covariate.
\begin{figure}[!ht]
     \centering
     \begin{subfigure}[b]{0.325\textwidth}
         \centering
         \includegraphics[trim=5pt 8pt 5pt 7pt, clip,width=\textwidth]{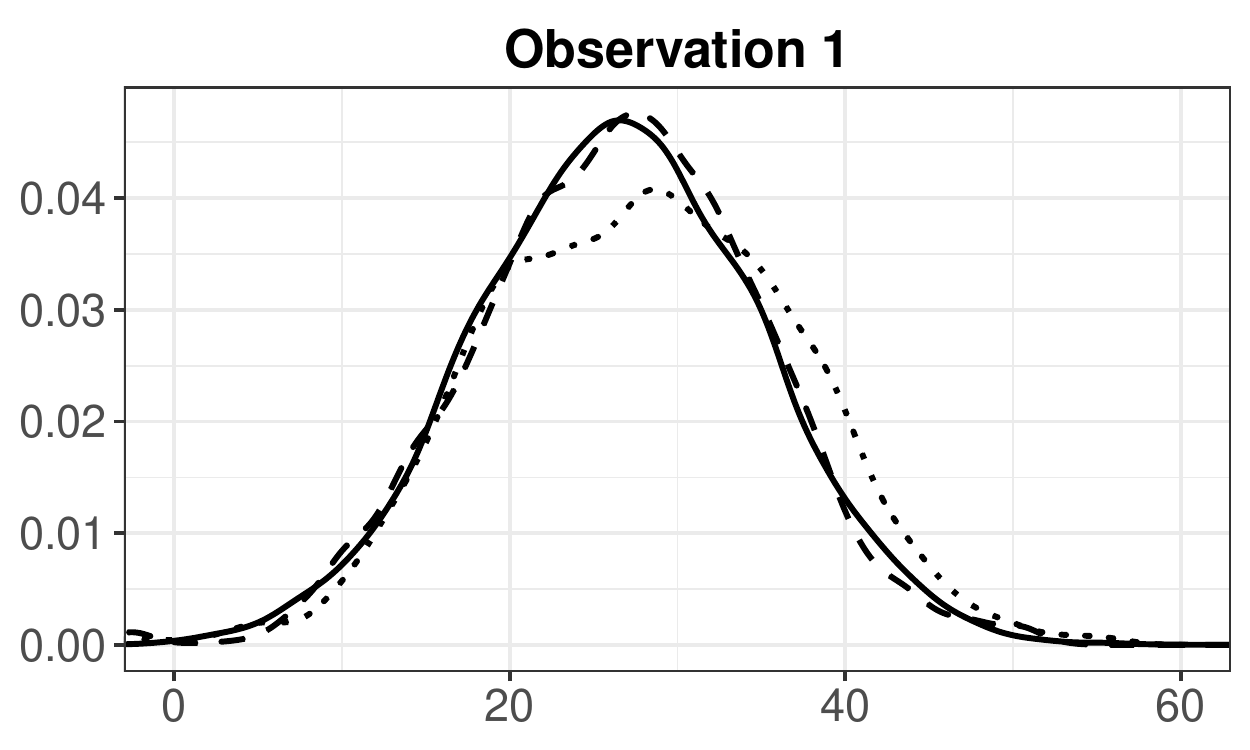}
     \end{subfigure}
     \begin{subfigure}[b]{0.325\textwidth}
         \centering
         \includegraphics[trim=5pt 8pt 5pt 7pt, clip,width=\textwidth]{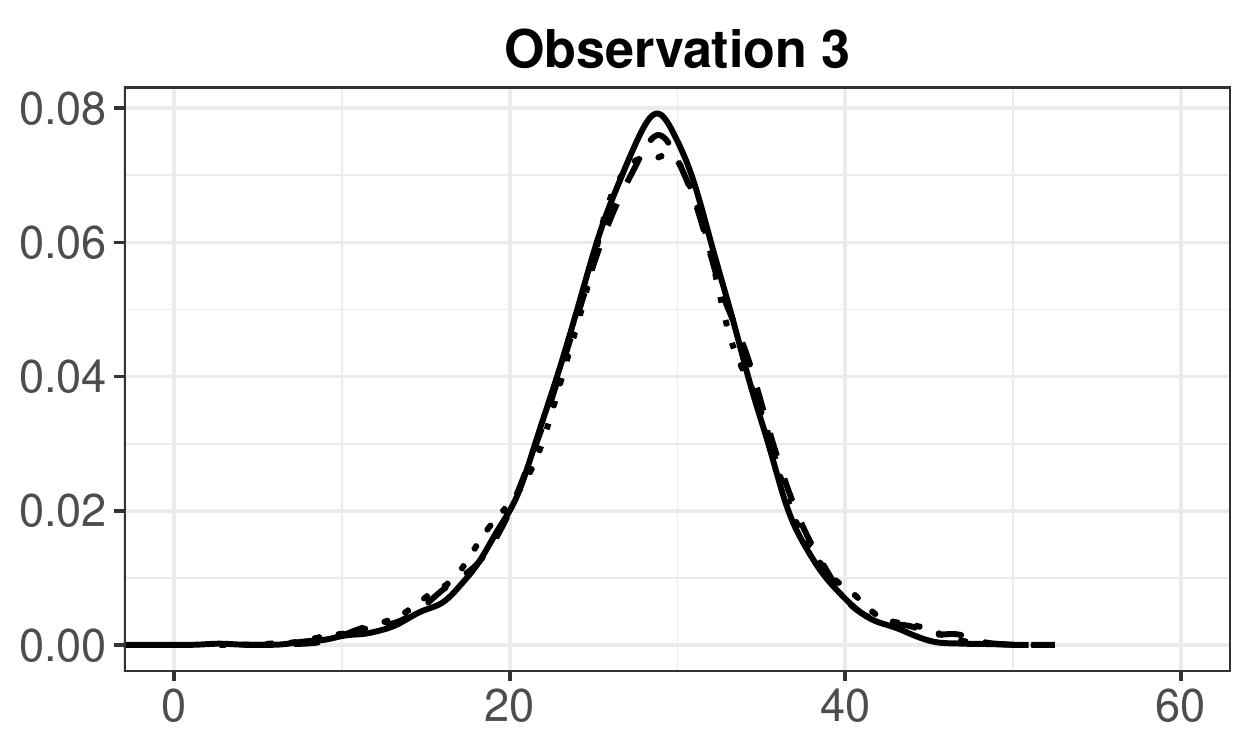}
     \end{subfigure}
     \begin{subfigure}[b]{0.325\textwidth}
         \centering
         \includegraphics[trim=5pt 8pt 5pt 7pt, clip,width=\textwidth]{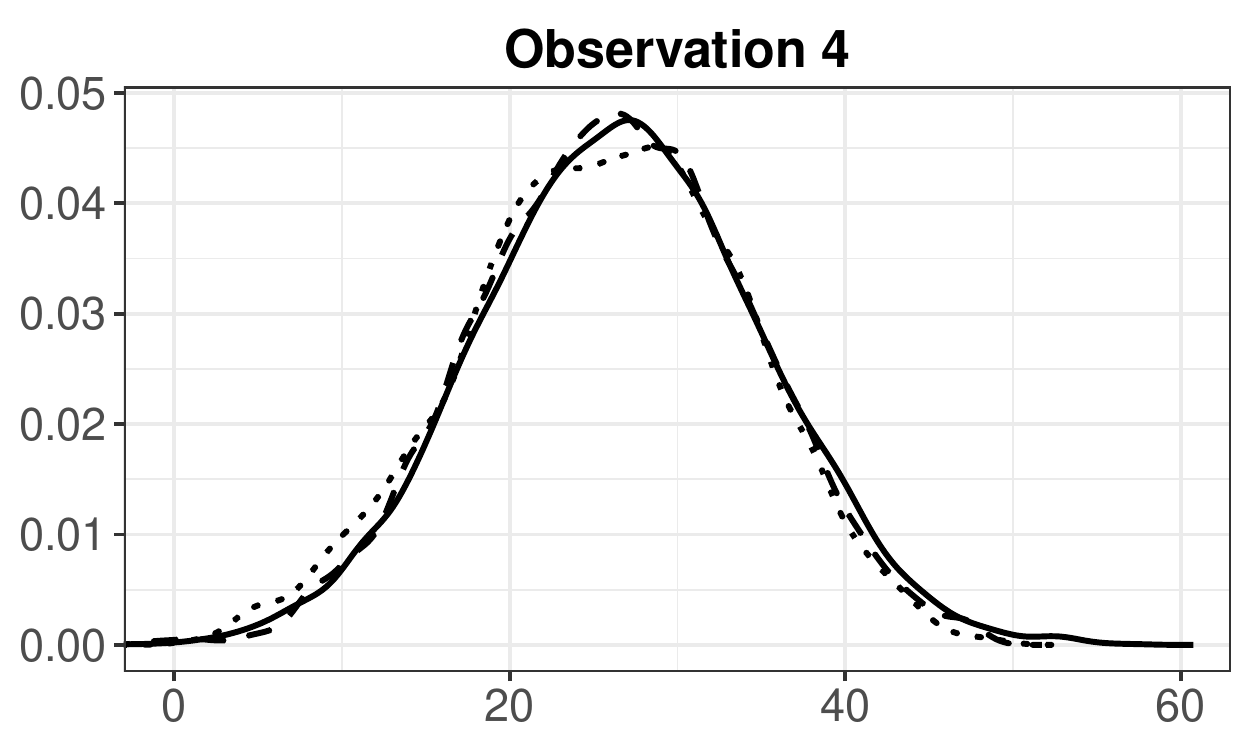}
     \end{subfigure}
     \begin{subfigure}[b]{0.325\textwidth}
         \centering
         \includegraphics[trim=5pt 8pt 5pt 7pt, clip,width=\textwidth]{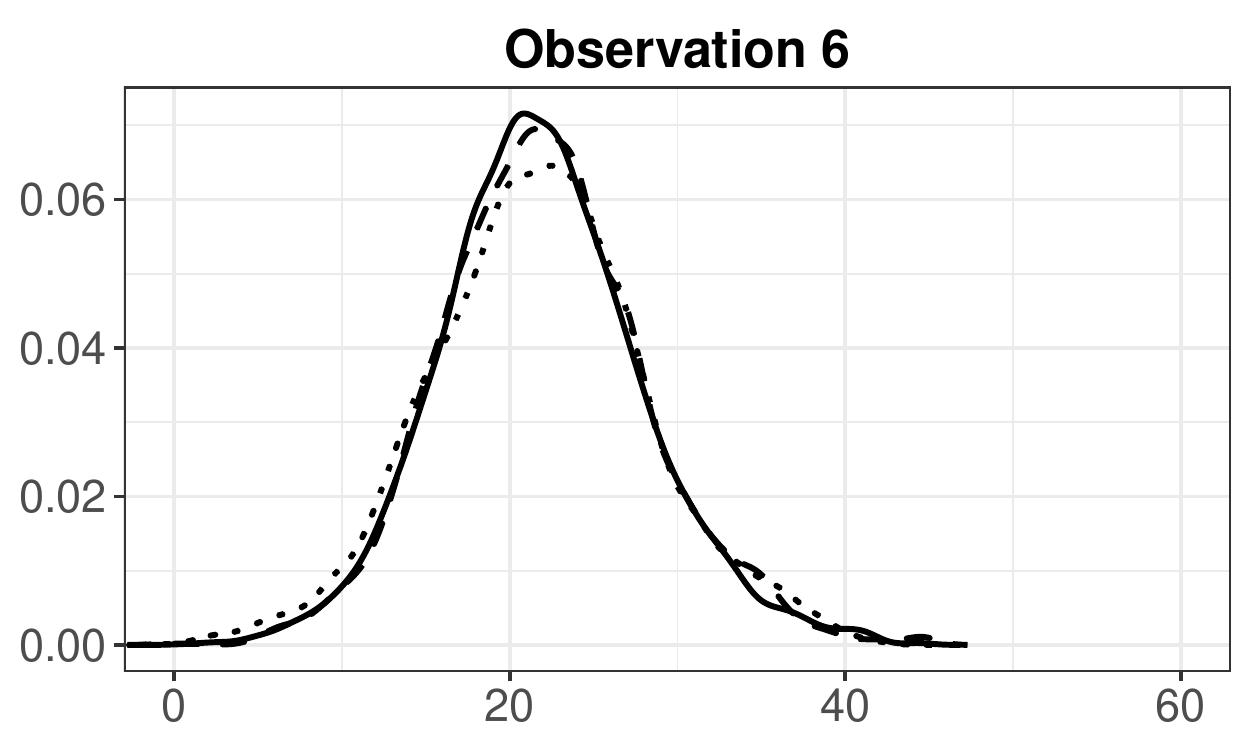}
     \end{subfigure}
     \begin{subfigure}[b]{0.325\textwidth}
         \centering
         \includegraphics[trim=5pt 8pt 5pt 7pt, clip,width=\textwidth]{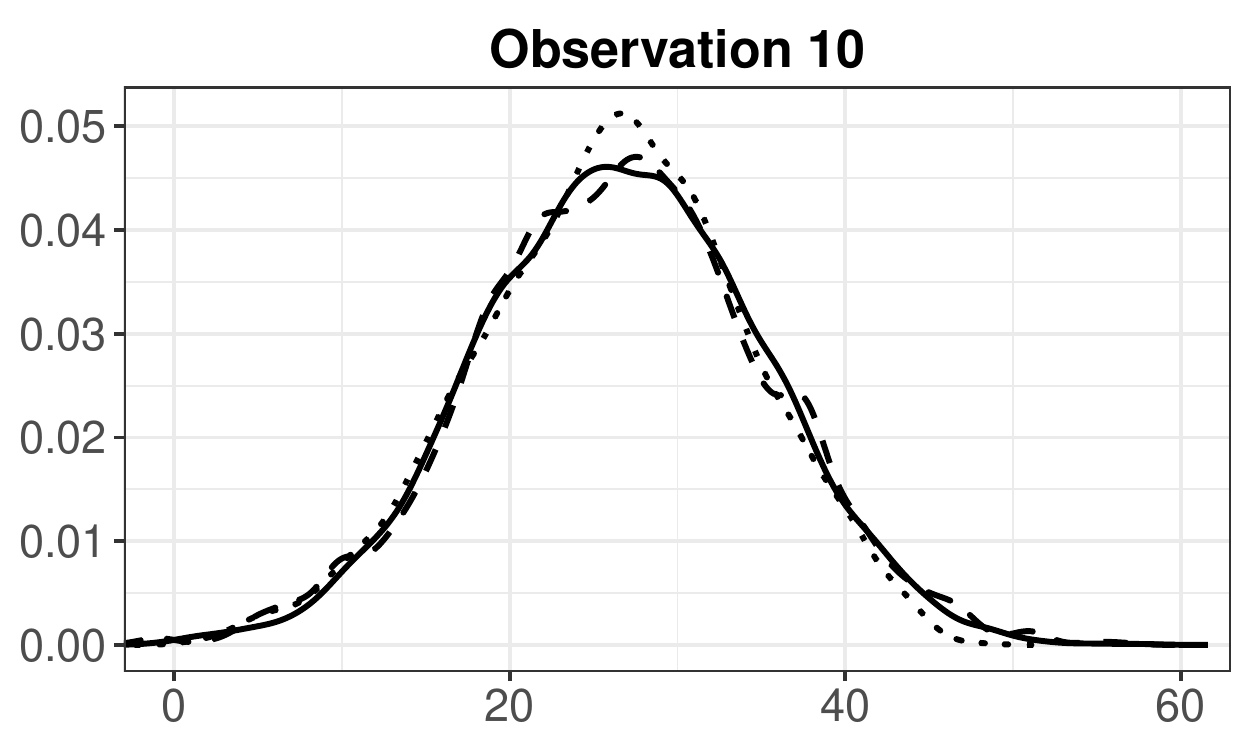}
     \end{subfigure}
     \begin{subfigure}[b]{0.325\textwidth}
         \centering
         \includegraphics[trim=5pt 8pt 5pt 7pt, clip,width=\textwidth]{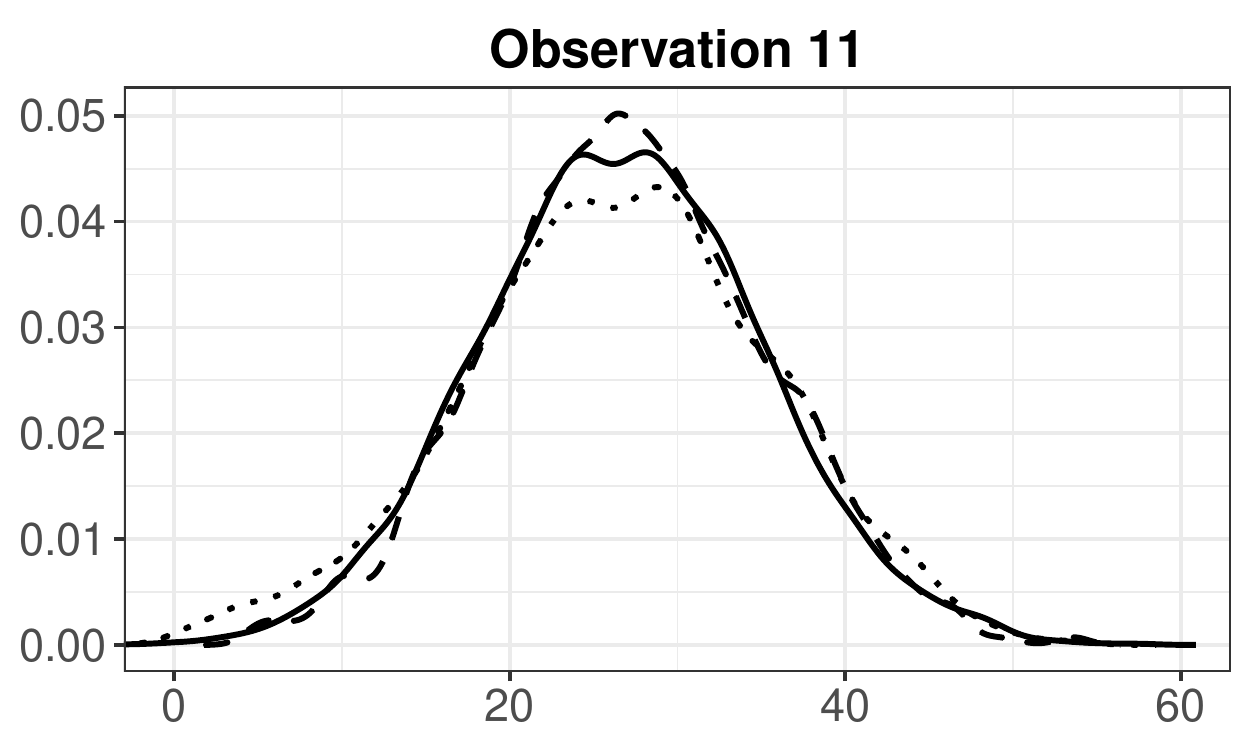}
     \end{subfigure}
     \begin{subfigure}[b]{0.325\textwidth}
         \centering
         \includegraphics[trim=5pt 8pt 5pt 7pt, clip,width=\textwidth]{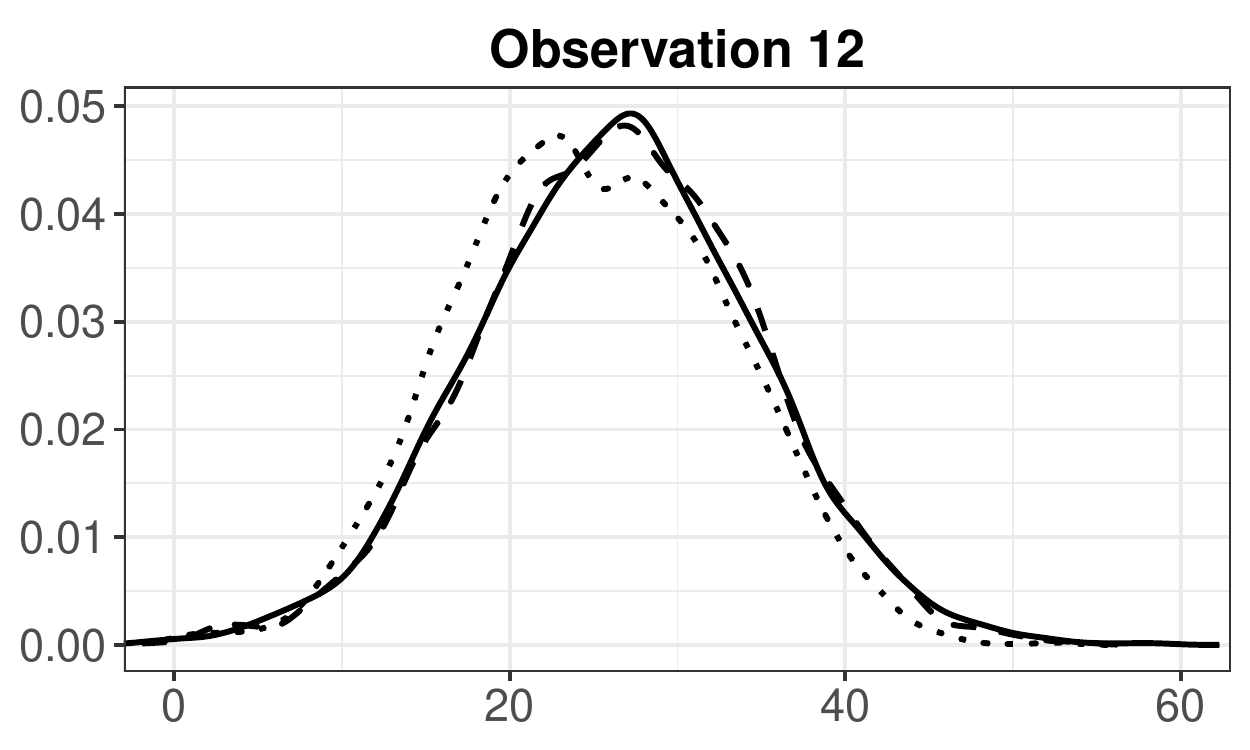}
     \end{subfigure}
     \begin{subfigure}[b]{0.325\textwidth}
         \centering
         \includegraphics[trim=5pt 8pt 5pt 7pt, clip,width=\textwidth]{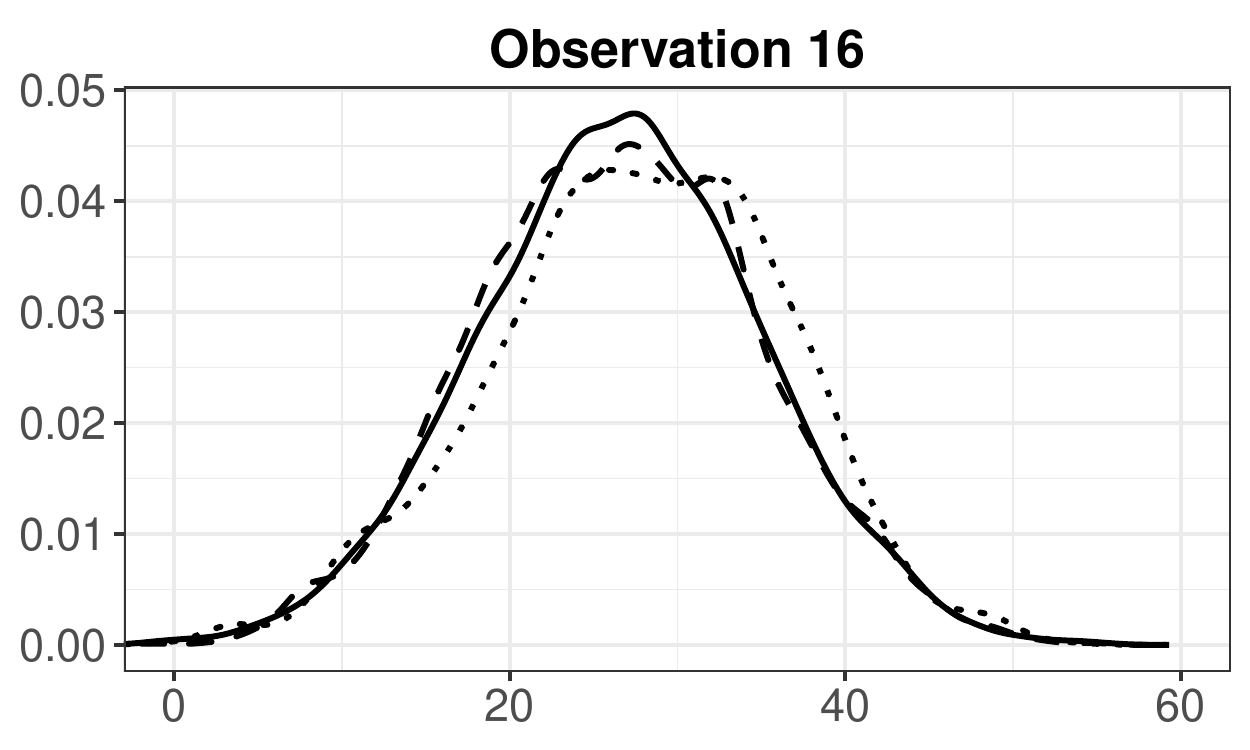}
     \end{subfigure}
     \begin{subfigure}[b]{0.325\textwidth}
         \centering
         \includegraphics[trim=5pt 8pt 5pt 7pt, clip,width=\textwidth]{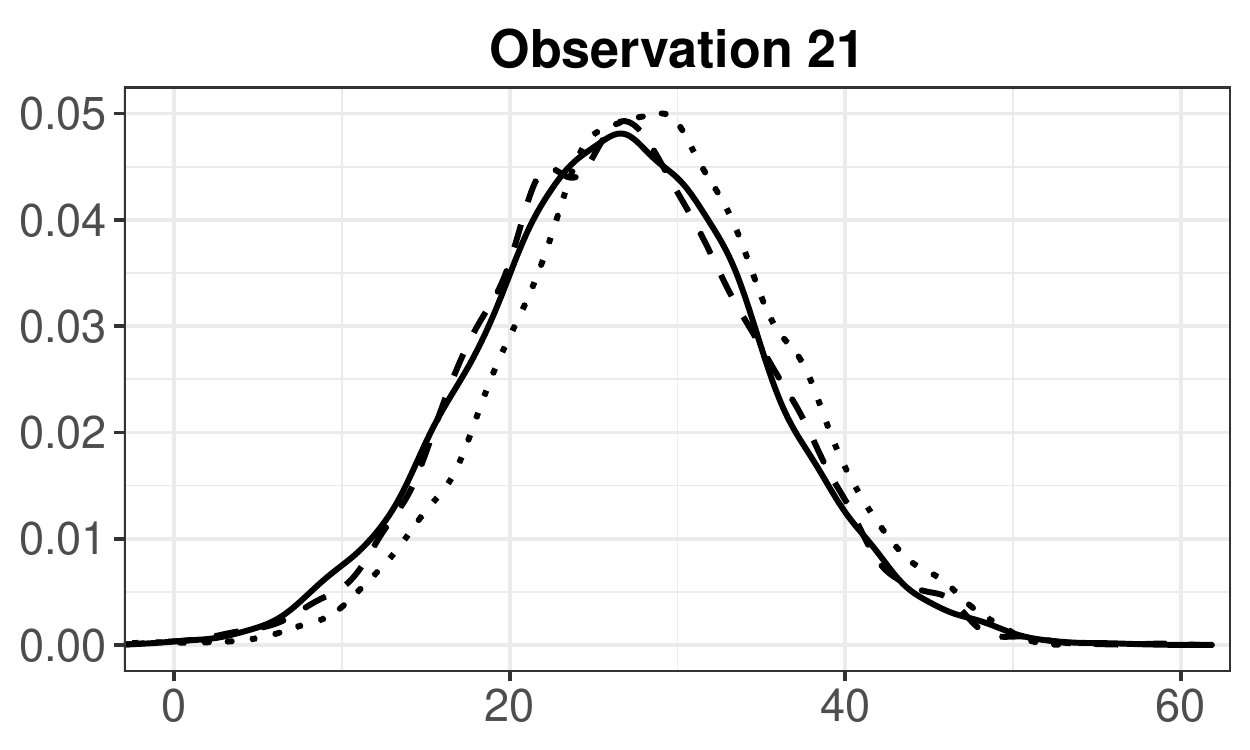}
     \end{subfigure}
    \caption{The Posterior marginals of the imputed missing values of the posterior distribution approximated using AMIS-INLA (\solid), IS-INLA (\dashed), MCMC-INLA (\dotted).}
        \label{fig:missing2}
\end{figure}

\begin{figure}[!ht]
     \centering
     \begin{subfigure}[b]{0.325\textwidth}
         \centering
         \includegraphics[trim=6pt 10pt 5pt 8pt,clip,width=\textwidth]{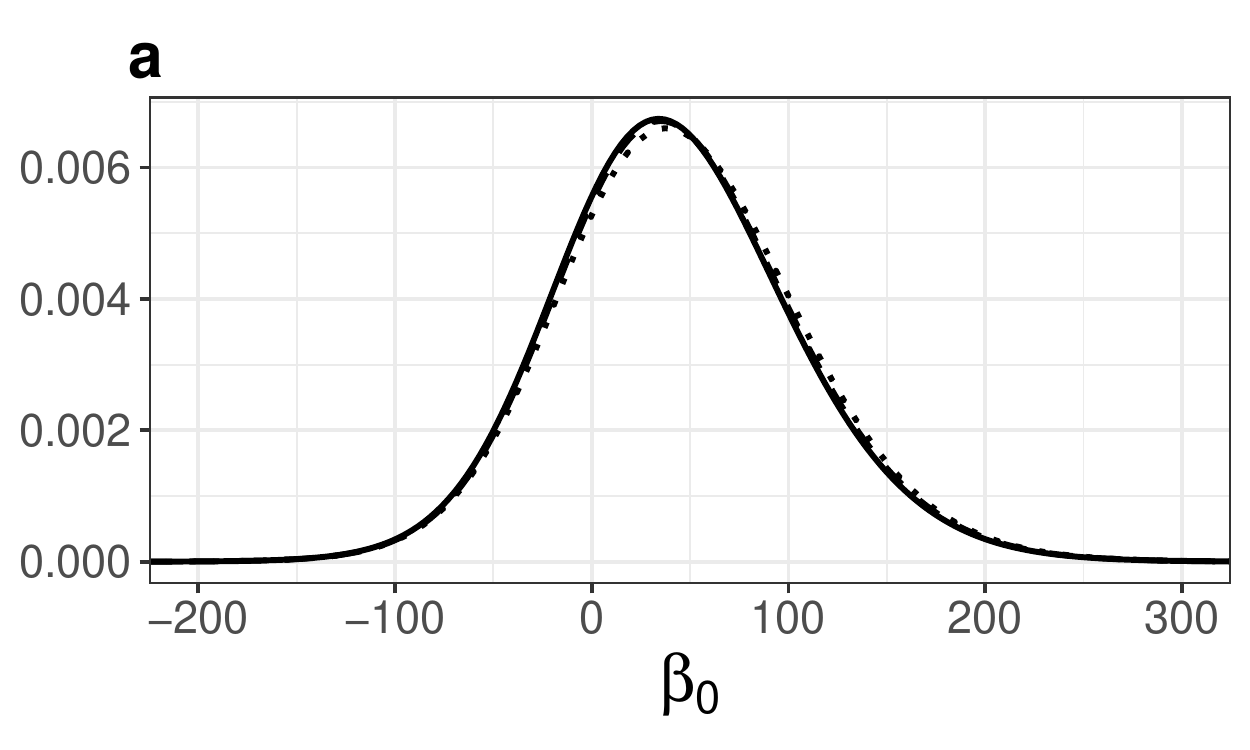}
     \end{subfigure}
     \begin{subfigure}[b]{0.325\textwidth}
         \centering
         \includegraphics[trim=6pt 10pt 5pt 8pt,clip,width=\textwidth]{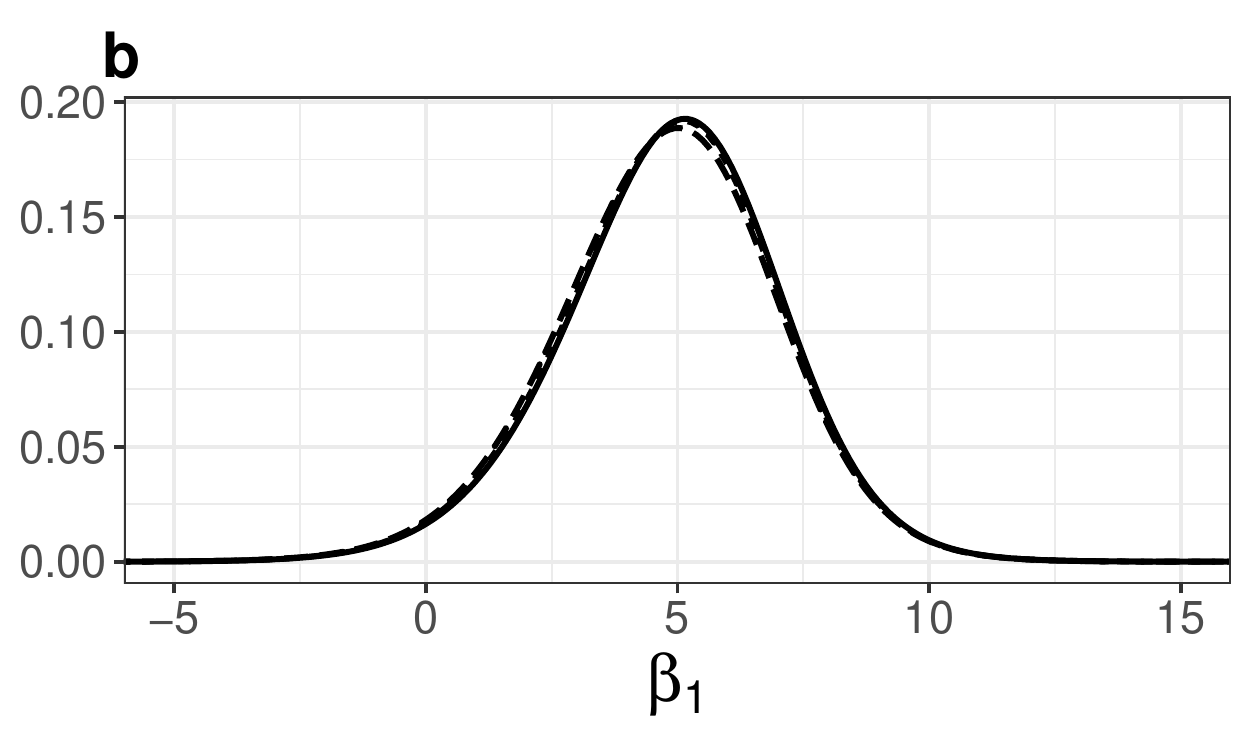}
     \end{subfigure}
     \begin{subfigure}[b]{0.325\textwidth}
         \centering
         \includegraphics[trim=6pt 10pt 5pt 8pt,clip,width=\textwidth]{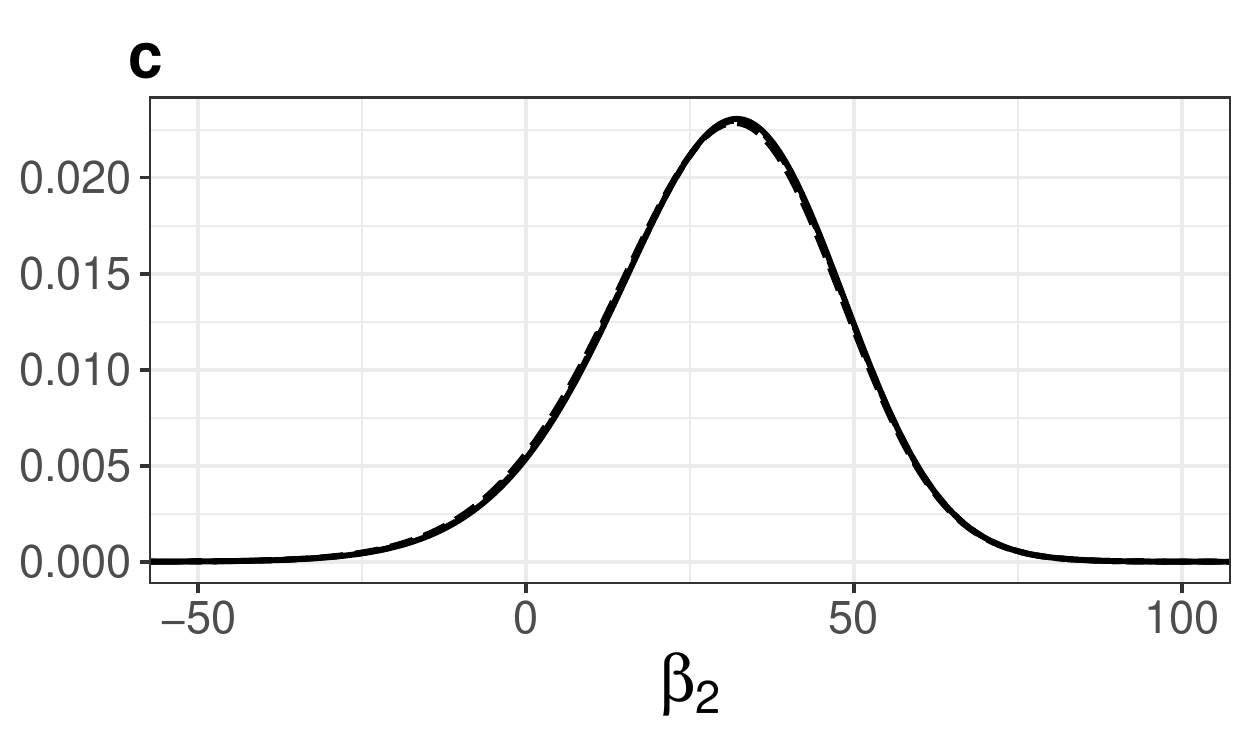}
     \end{subfigure}
      \begin{subfigure}[b]{0.325\textwidth}
         \centering
         \includegraphics[trim=6pt 10pt 5pt 8pt,clip,width=\textwidth]{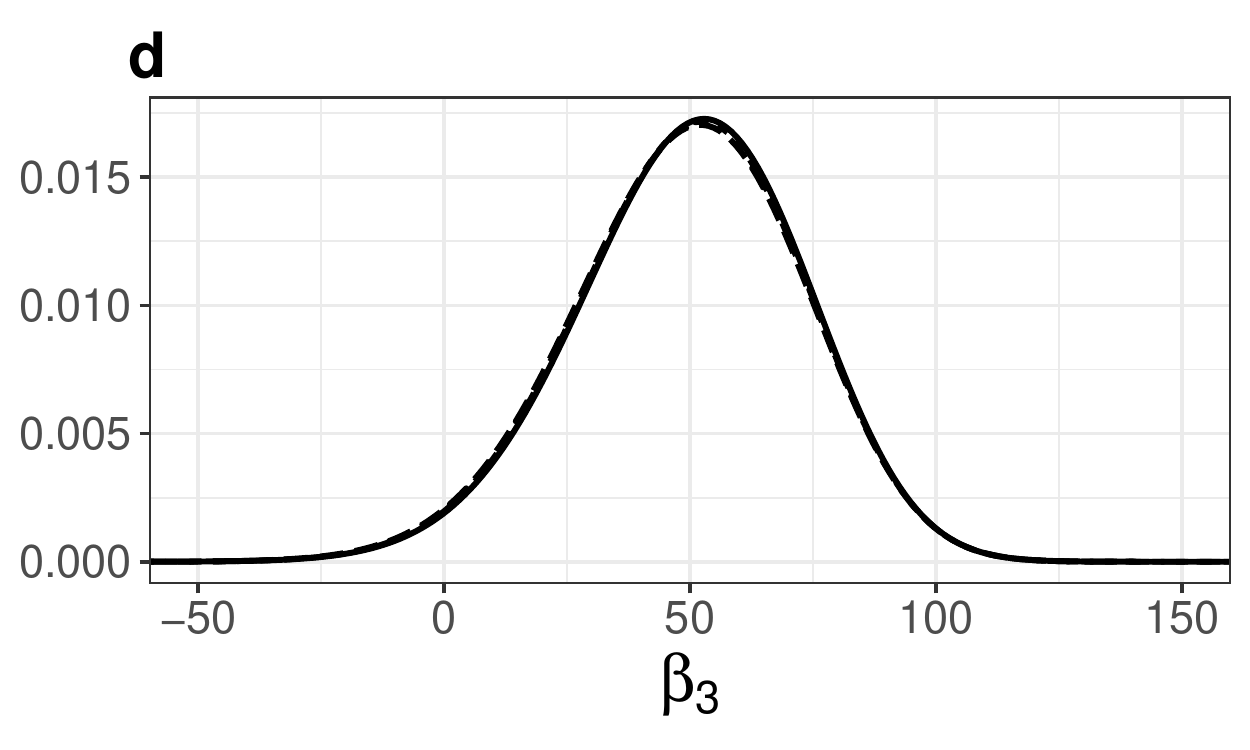}
     \end{subfigure}
          \begin{subfigure}[b]{0.325\textwidth}
         \centering
         \includegraphics[trim=6pt 10pt 5pt 8pt,clip,width=\textwidth]{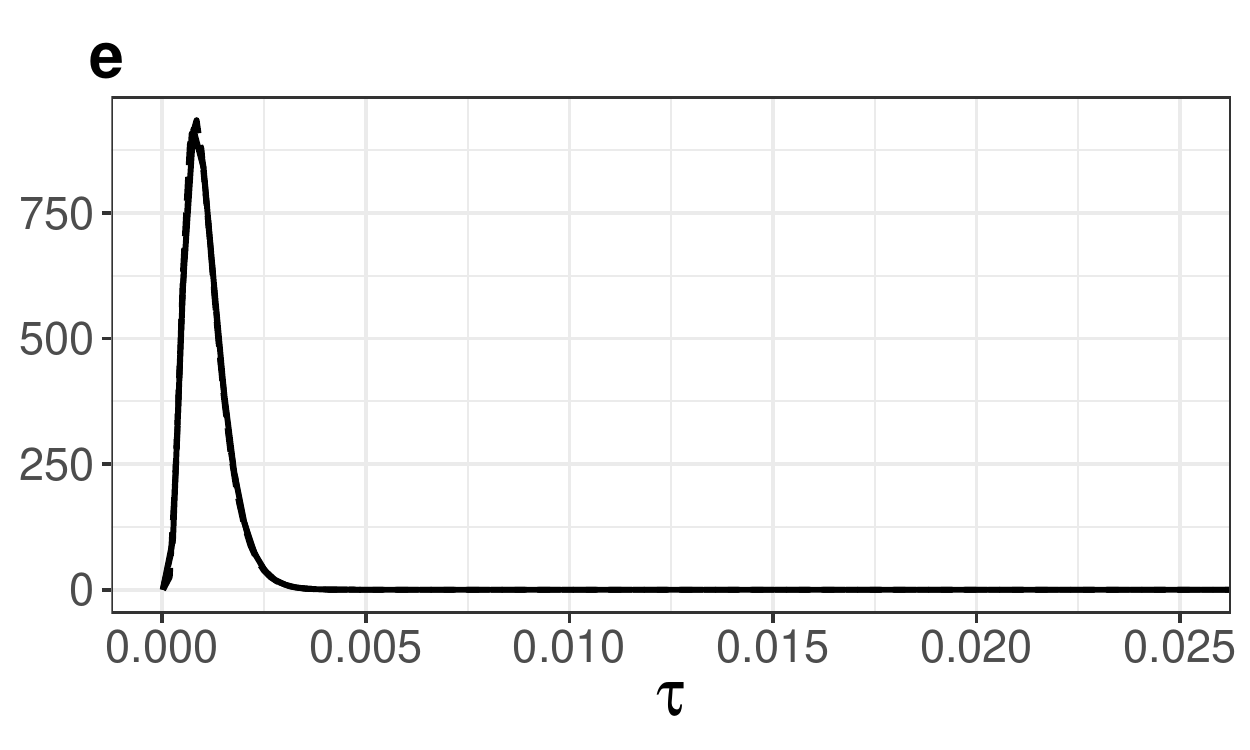}
     \end{subfigure}
     \begin{subfigure}[b]{0.325\textwidth}
         \centering
         \includegraphics[trim=6pt 10pt 5pt 8pt,clip,width=\textwidth]{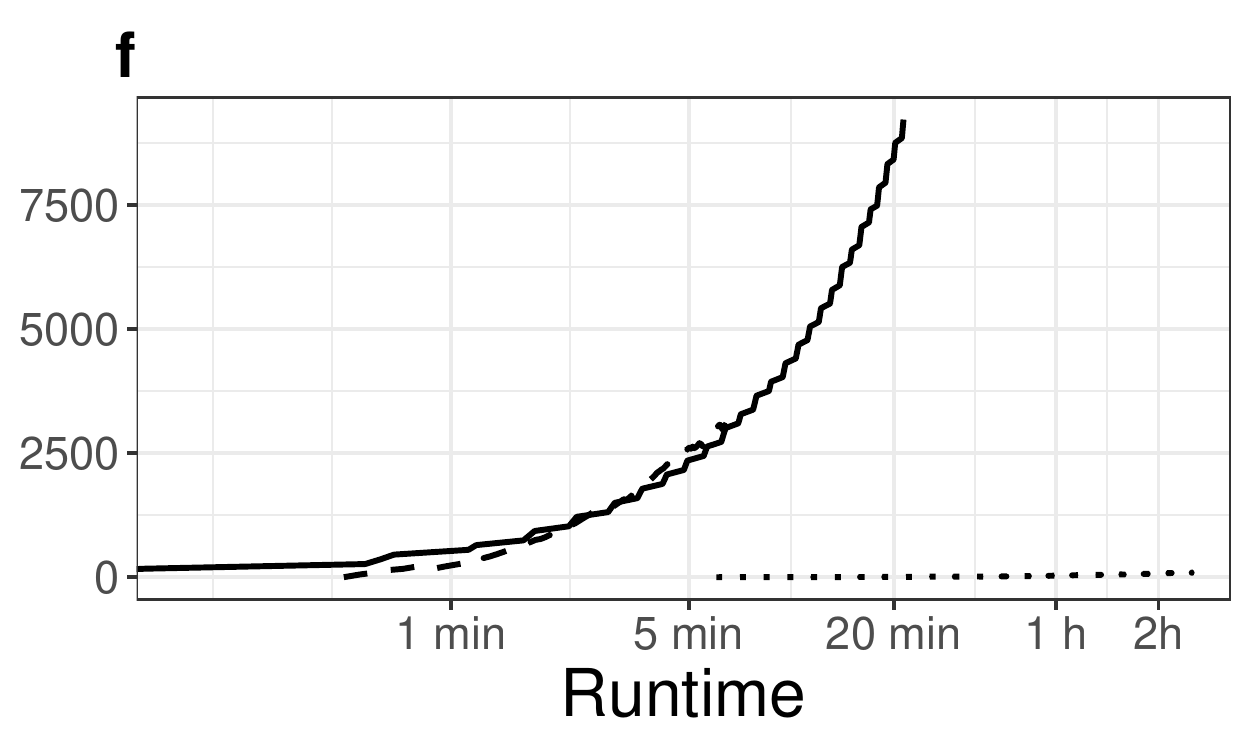}
     \end{subfigure}
    \caption{The posterior marginals of the coefficients in the model with missing covariates (a-e), and the running effective sample size (f) obtained using AMIS with INLA (\solid), IS with INLA (\dashed) and MCMC with INLA (\dotted). The runtimes is presented in logarithmic scale.}
        \label{fig:missing1}
\end{figure}

Posterior marginals of the imputed missing covariates are shown in Figure~\ref{fig:missing2}, while posterior marginal for the model parameters are shown in Figure~\ref{fig:missing2}a-e). All  approaches  give similar  estimates. The difference can be appreciated  by looking at the running effective sample size  in Figure~\ref{fig:missing1}.
MCMC-INLA run for more than 9 hours, with an effective samples size of 1072, that is 0.03 effective samples per second. Both IS-INLA and AMIS-INLA appear to be much more efficient with respectively 8.02 and 7.19 effective samples per second.

Per-variable values of the sample size $n_e(h)$ range from 2579.964 (for observation 10) and 3105.985 (observation 11) for IS-INLA, and between 8293.174 (observation 6)  and 8468.784 (observation 10) for AMIS-INLA. All the probability plots look very good, with lines very close to the identity line, and they are not shown here but are avialable in the Suplementary Materials.

%% file: pqr.tex
\subsection{Model-based Bayesian quantile regression}
\label{subsec:pqr}

Quantile regression is used to understand the relationship between the quantiles of the response and some covariates and was introduced by \citet{koenker_regression_1978}. The frequentist approach to quantile regression is well developed and relies on minimizing a loss function. In the Bayesian framework, a common approach to quantile regression is to employ the asymmetric Laplace distribution  as likelihood model \citep{yu_bayesian_2001}.
Such likelihood is a mere working likelihood and does not describe the data generation process. 
We follow here \citet{noufaily_parametric_2013}  and \citet{padellini_model-aware_2019} that propose instead a  parametric approach to quantile regression. \citet{padellini_model-aware_2019} work with Poisson data and propose to create a direct link between the quantiles of the response and the linear predictor. 
\citet{noufaily_parametric_2013} show that by modeling all likelihood parameters as a function of covariates, interesting shapes are found in the quantile curves. An important advantage of such approach is that quantile curves cannot cross, a major issue covered in many studies \citep[see, for example,][]{rodrigues_regression_2017}. 

Here we follow \cite{noufaily_parametric_2013}  and present one example of semi-parametric quantile regression for Gaussian data.  Note however that any other distribution can be treated in the same way.

\begin{figure}[!ht]
    \centering
    \includegraphics[width=0.6\textwidth]{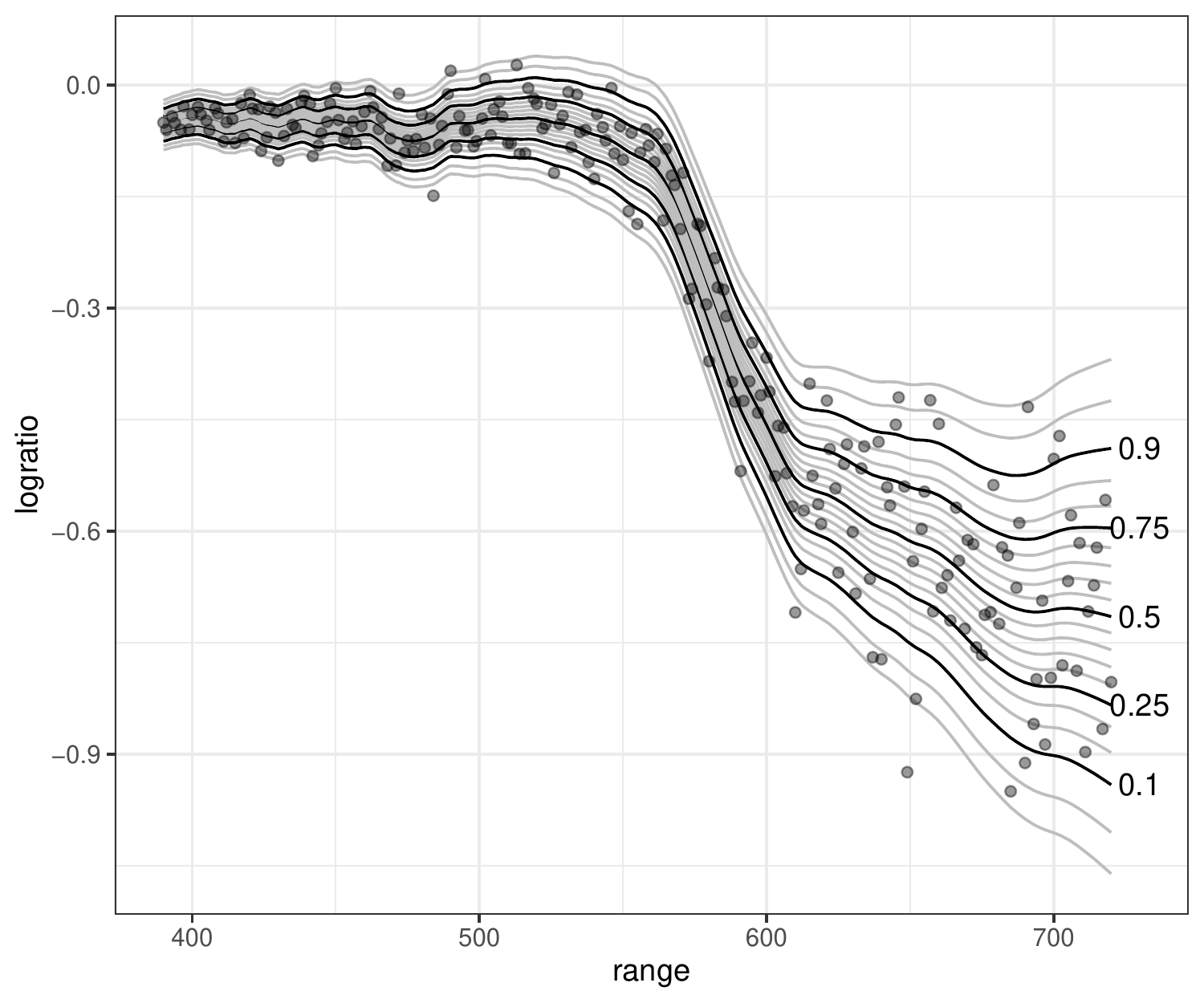}
    \caption{Estimated quantile curves of the second order random walk model on the LIDAR dataset obtained using the AMIS-INLA algorithm. The light grey lines are quantile curves in the range $p\in(0.025,0.975)$.}
    \label{fig:pqr_lidar}
\end{figure}

We  consider the LIDAR dataset \citep{sigrist_air_1994} available in the {\textsf{\textbf{R}}} library {\tt SemiPar} \citep{SemiPar}. It contains $n=221$ observations of two variables; the logarithm of the ratio of light received by two lasers, which we consider the response, and  the distance the light has traveled before it is reflected back to its source. The data are plotted in Figure~\ref{fig:pqr_lidar}, and it is clear  that both the mean and dispersion of the response variable depend on the observed value of the covariate.
Let $\bm{y} = (y_1,\dots,y_n)$ be the vector of observations. We assume
\begin{equation}
    y_i\sim\mathcal{N}(\mu_i,\sigma_i^2).
    \label{eq:lidar_lik}
\end{equation}
\noindent
Moreover we let:
\begin{equation}
    \mu_i = \mu_0 + f(x_i)
    \label{eq:rw2_linpred}
\end{equation}
where $\mu_0$ is an intercept and $f(\cdot)$ is a smooth effect of the covariate. As prior model for $f(\cdot)$ we assume a random walk of the second order \citep[see][for details]{RueHeld:2005} with precision parameter $\tau_{_\mathrm{RW}}$. Furthermore, the precision of the distribution of $y_i$ is modelled as $\log(\tau_i^2) = \alpha + \beta\cdot x_i$ and, thus, the log standard deviation in Equation~\eqref{eq:lidar_lik} is
\begin{equation}
    \log(\sigma_i) = -\frac{1}{2}(\alpha + \beta\cdot x_i)
        \label{eq:lidar_sd}
\end{equation}
The model is completed by assigning vague Gaussian priors to $\mu_0,\alpha $ and $\beta$ and Gamma with parameters 1 and 0.00005 for the precision parameter $\tau$. The $p$ quantile $y_p$ for the model in Equations \ref{eq:lidar_lik}-\ref{eq:lidar_sd} can be computed as:
\begin{equation}
    y_p(x) = \mu_0 + f(x) + \sqrt{\exp\left(-\frac{1}{2}(\alpha + \beta\cdot x)\right)} y^*_p,
    \label{eq:pqr_gaussian}
\end{equation}
where $y^*_p,$ is the $p$ quantile of the standard Gaussian distribution.

The model in Equations~(\ref{eq:lidar_lik}-\ref{eq:lidar_sd}) cannot be fit in R-INLA unless we condition on the parameter vector $\bm{z}_c = (\alpha,\beta)$. We therefore use MCMC-INLA, IS-INLA and AMIS-INLA to fit the model to the data.  Both MCMC-INLA and IS-INLA required a good deal of tweaking  in order to obtain a sufficient performance. AMIS-INLA instead managed to adapt the proposal automatically. The starting proposal used here is a Student-$t$ distribution with $\nu = 3$, $\bm{\mu}_0 = 0$ and $\bm{\Sigma}_0 = 10 \cdot \mathbf{I}$. 

\begin{figure}[!ht]
     \centering
     \begin{subfigure}[b]{0.49\textwidth}
         \centering
         \includegraphics[trim=6pt 9pt 5pt 5pt, clip,width=\textwidth]{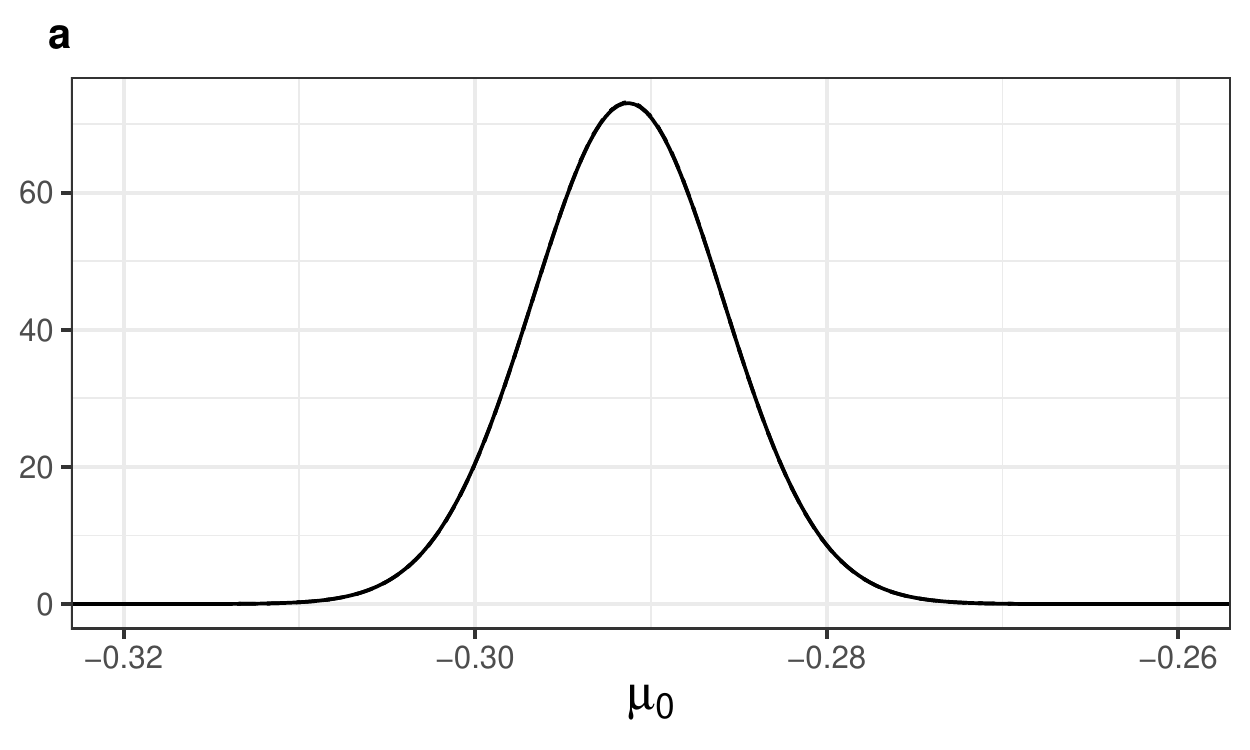}
     \end{subfigure}
     \begin{subfigure}[b]{0.49\textwidth}
         \centering
         \includegraphics[trim=6pt 9pt 5pt 5pt, clip,width=\textwidth]{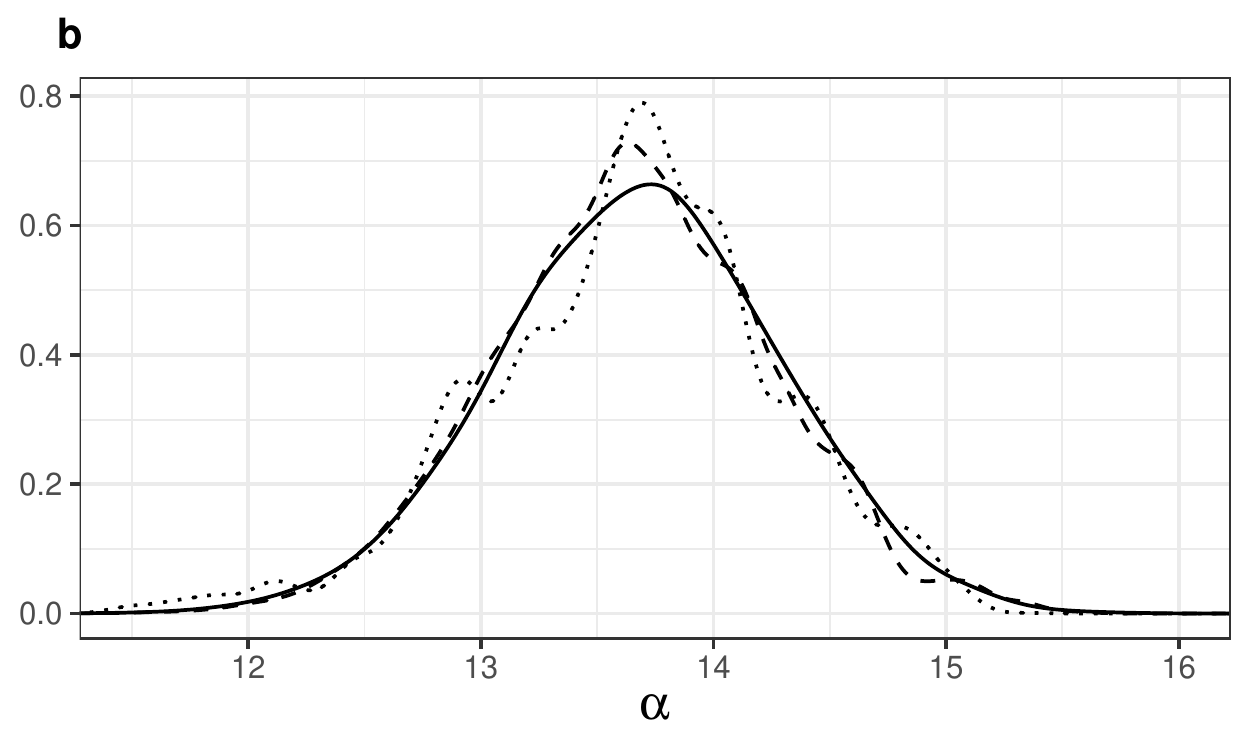}
     \end{subfigure}
     \begin{subfigure}[b]{0.49\textwidth}
         \centering
         \includegraphics[trim=6pt 9pt 5pt 5pt, clip,width=\textwidth]{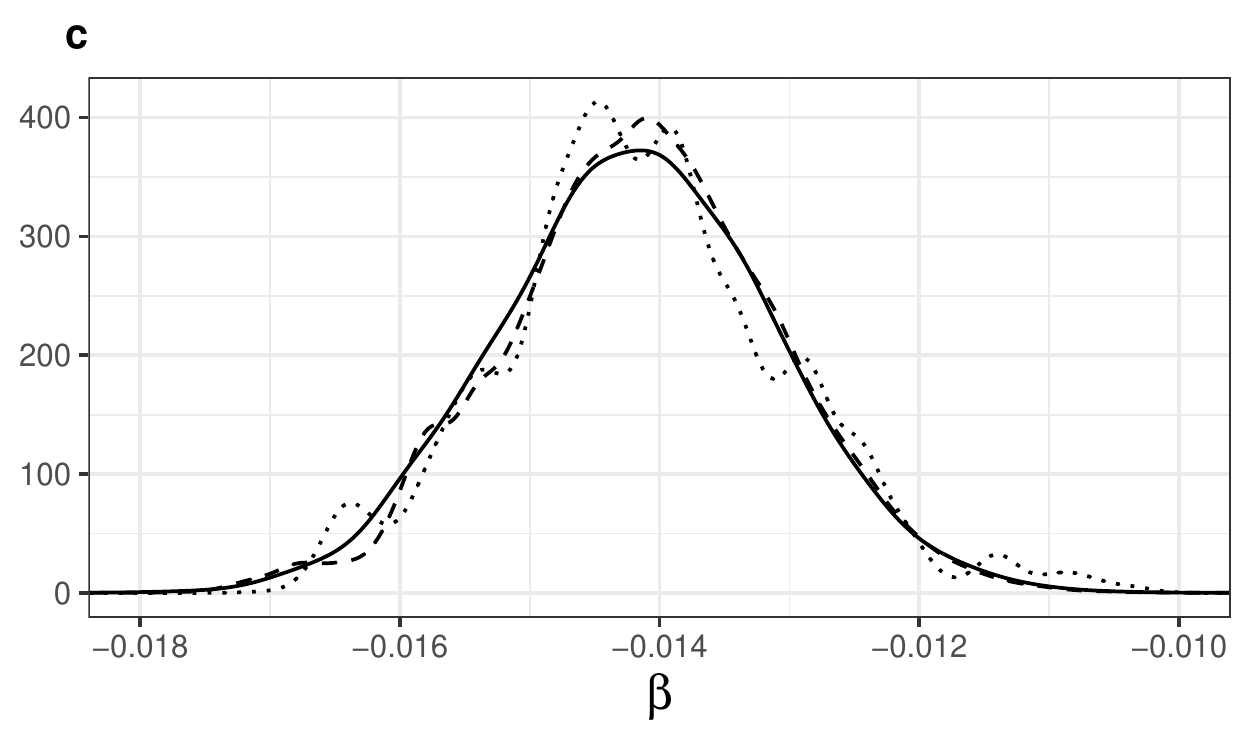}
     \end{subfigure}
      \begin{subfigure}[b]{0.49\textwidth}
         \centering
         \includegraphics[trim=6pt 9pt 5pt 5pt, clip,width=\textwidth]{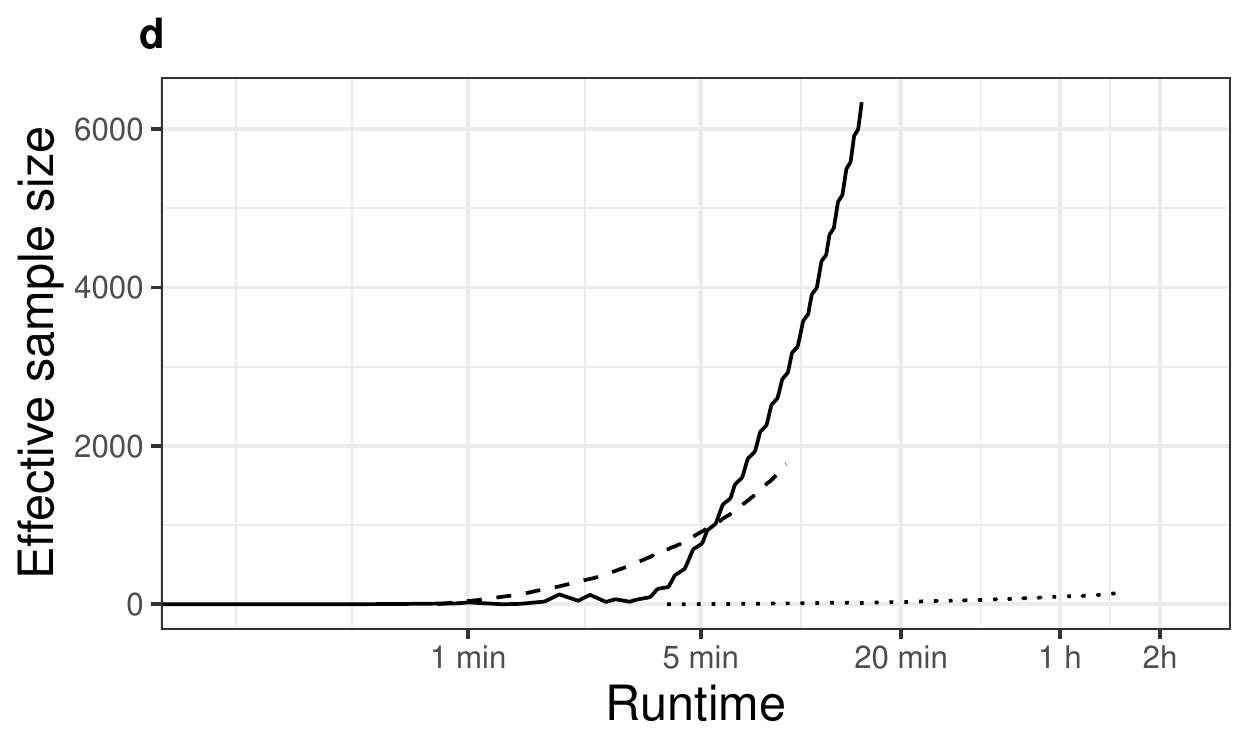}
     \end{subfigure}
    \caption{Posterior marginals of $\bm{z}=(\mu_0,\alpha,\beta)$ (a-c), and the running effective sample size (d) in the second order random walk model for LIDAR data approximated with AMIS-INLA (\solid), IS-INLA (\dashed), and MCMC-INLA (\dotted).}
    \label{fig:pqr_lidar_uni}
\end{figure}

\begin{figure}[!ht]
     \centering
     \begin{subfigure}[b]{0.49\textwidth}
         \centering
         \includegraphics[width=\textwidth]{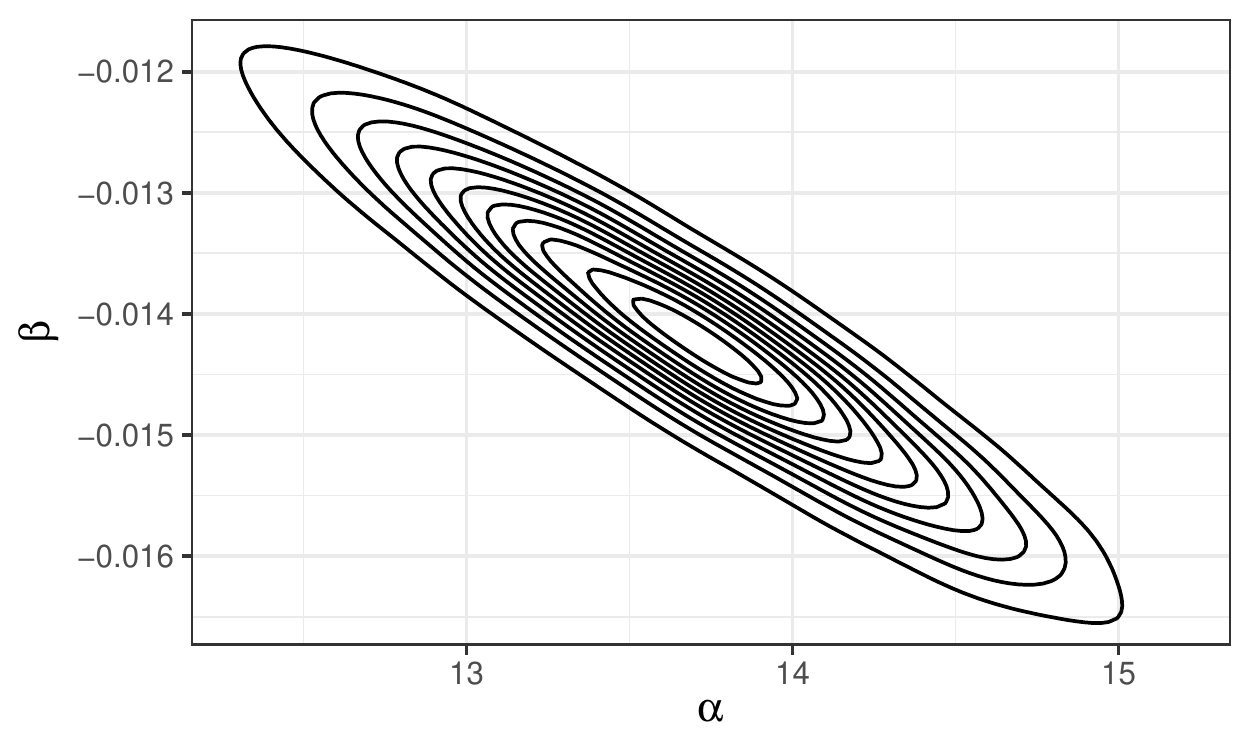}
     \end{subfigure}
     \begin{subfigure}[b]{0.49\textwidth}
         \centering
         \includegraphics[width=\textwidth]{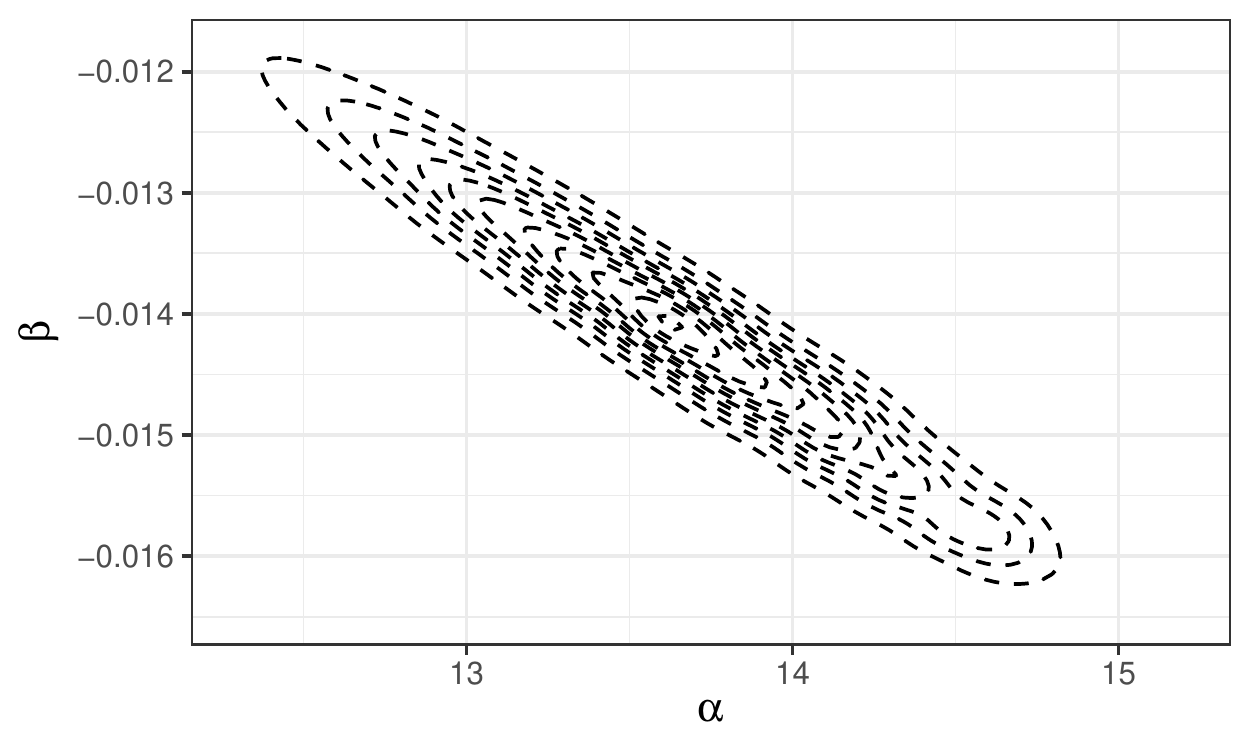}
     \end{subfigure}
      \begin{subfigure}[b]{0.49\textwidth}
         \centering
         \includegraphics[width=\textwidth]{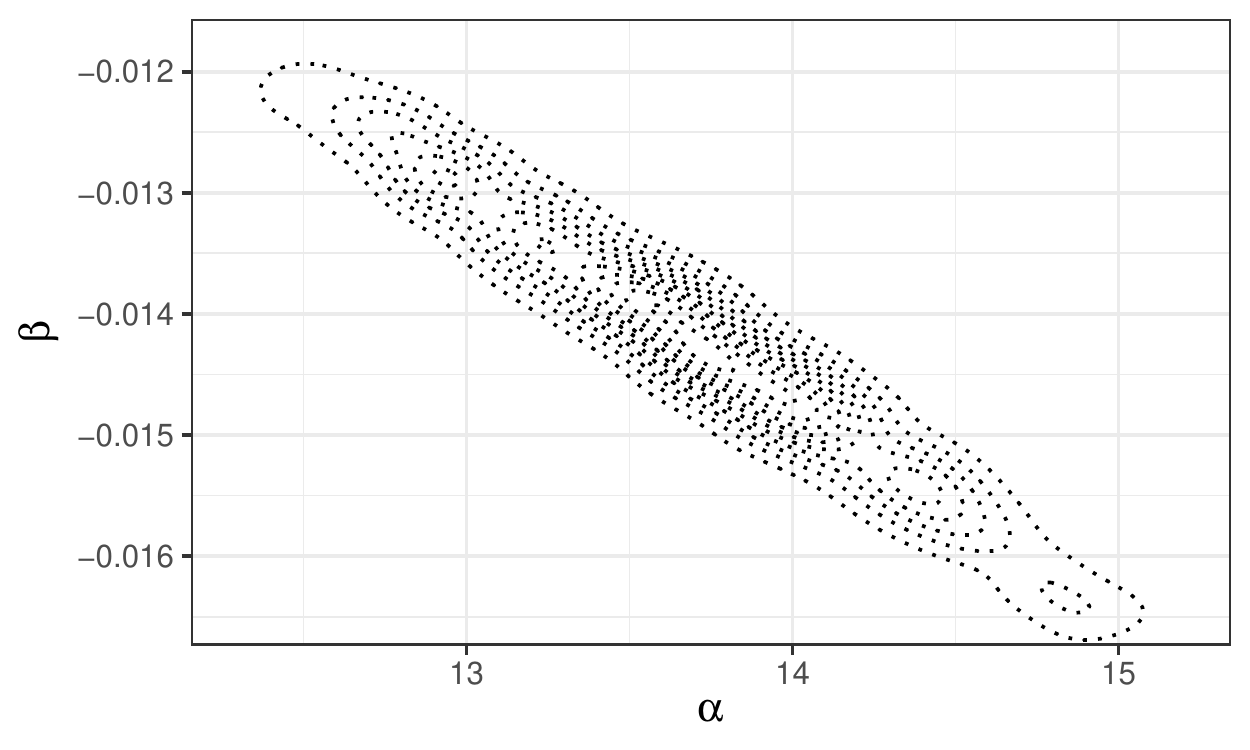}
     \end{subfigure}
    \caption{Joint posterior distribution of $\bm{z}_c=(\alpha,\beta)$ in the random walk model for LIDAR data approximated with AMIS-INLA (\solid), IS-INLA (\dashed), and MCMC-INLA (\dotted).}
    \label{fig:pqr_lidar_joint}
\end{figure}

Figure~\ref{fig:pqr_lidar_uni} shows the posterior distributions for the parameters $\mu_0, \alpha$    and  $\beta$ obtained with the three methods together with the respective running  effective sample size, while Figure~\ref{fig:pqr_lidar_joint} shows the joint posterior density for $\alpha$ and $\beta$. Posterior estimates are very similar even if Monte Carlo error is clealy visible in the MCMC-INLA estimates. AMIS-INLA has clearly managed to produce more effective samples than the other two methods.

This, together with the fact that AMIS-INLA did not require manual tuning shows that in this situation, the adaptive nature of this algorithm is clearly an advantage. 


Finally, Figure~\ref{fig:pqr_lidar} shows the estimated quantile curves obtained with AMIS-INLA together with the observed data. 

Regarding IS diagnostics, the values of the effecive sample size $n_e(h)$ for parameters $\alpha$ and $\beta$ are 1657.741 and 1582.558 for IS-INLA and 6312.414 and 6298.910 for AMIS-INLA. Hence, this is in line with the results already commented above. The probability plots also look much better for AMIS-INLA than IS-INLA (see Suplementary Materials).